\documentclass{aa}  
\usepackage{graphicx}
\usepackage{txfonts}
\usepackage{natbib}

\def\aaps{A\&AS}
\def\aap{A\&A}
\def\apj{ApJ}
\def\apjs{ApJS}
\def\aj{AJ}

\def\ha{H\,$\alpha$}
\def \hi {H\,{\sc i~}}

\def\NH{$N_{\rm HI}$}
\def\kms{km\,s$^{-1}$}
\def\deg{\hbox{$^\circ$}}
\def\arcmin{\hbox{$^\prime$}}

\def\fdg{\hbox{$.\!\!^\circ$}}

\begin{document}
   \title{Global properties of the HI high velocity sky}

   \subtitle{a statistical investigation based on the LAB survey}

   \author{P.\,M.\,W. Kalberla\inst{1} \& U. Haud\inst{2}  }

   \institute{ Argelander-Institut f\"ur Astronomie, Universit\"at 
Bonn\thanks{Founded by merging of the
Sternwarte, Radioastronomisches Institut and Institut f\"ur Astrophysik
und Extraterrestrische Forschung der Universit\"at Bonn},
              Auf dem H\"ugel 71, 53121 Bonn, Germany\\
              \email{pkalberla@astro.uni-bonn.de}
           \and
              Tartu Observatory, 61602 Toravere, Estonia\\
              \email{urmas@aai.ee} }
      
   \authorrunning{P.\,M.\,W. Kalberla \& U. Haud } 

   \titlerunning{Global properties of the HI high velocity sky}

   \offprints{P.\,M.\,W. Kalberla}

   \date{Received December 22 2005/ Accepted  April 10 2006 }

 
  \abstract 
  {Since 1973 it is known that some \hi high velocity clouds (HVCs) have
  a core-envelope structure. Recent observations of compact HVCs 
  confirm this but more general investigations have been missing so far.}
  {We study the properties of all major HVC complexes from a sample
  compiled 1991 by Wakker \& van Woerden (WvW). }
  {We use the Leiden/Argentine/Bonn all sky 21-cm line survey and
  decompose the profiles into Gaussian components.}
  {We find the WvW line widths and column densities underestimated by
  $\sim 40$\%. In 1991 these line widths could not be measured directly
  but had to be estimated with the help of higher-resolution data. We
  find a well defined multi-component structure for most of the HVC
  complexes. The cold HVC phase has lines with typical velocity
  dispersions $\sigma = 3 $ \kms and exists only within more extended
  broad line regions, typically with $\sigma = 12 $ \kms.  The motions
  of the cores relative to the envelopes are characterized by Mach
  numbers $M = (v_{core}-v_{envelope})/\sigma_{envelope} \sim 1.5$. The
  center velocities of the cores within a HVC complex have typical
  dispersions of 20 \kms. Remarkable is the well defined two-component
  structure for some prominent HVC complexes in the outskirts of the
  Milky Way: Complex H lies approximately in the galactic plane, and the
  most plausible distance estimate of $R \sim 33$ kpc places it at the
  edge of the disk. The Magellanic Stream and the Leading Arm (complex
  EP) reach higher latitudes and are probably more distant, $R \sim 50$
  kpc. There might be some indications for an interaction between HVCs
  and disk gas at intermediate velocities. This is possible for complex
  H, M, C, WB, WD, WE, WC, R, G, GCP, and OA, but not for complex A, MS,
  ACVHV, EN, WA, and P.  }
  {The line widths, determined by us, imply that estimates of HVC masses,
  as far as derived from the WvW database, need to be scaled up by a
  factor 1.4. Correspondingly, guesses for the external pressure of a
  confining coronal gas need to be revised upward by a factor of 2. The
  HVC multi-phase structure implies in general that currently the halo
  pressure is significantly underestimated. In consequence, the HVC
  multi-phase structure may indicate that most of the complexes are
  circum-galactic. HVCs have turbulent energy densities which are an
  order of magnitude larger than that of comparable clumps in the
  Galactic disk.}
  \maketitle
%

\section{Introduction}

Forty years after the discovery of high velocity clouds (HVCs) outside
the disk of the Milky Way \citep{Muller} the nature of these 
distinguished objects has still remained a mystery. Large quantities of
data have been collected since then and numerous explanations for their
origins have been proposed, we refer to the review by \citet{WvW97} and
volume 312 of the Astrophysics and Space Science Library, dedicated to
the HVC phenomenon \citep{HVCbook}. An outstanding contribution and one
of the roots for our present understanding of the structure and
distribution of HVCs on large scales is the catalog of individual HVCs
by \citet{WvW} (hereafter WvW).

Our contribution to this topic, presented here, is based on the
Leiden/Argentine/Bonn (LAB) \hi line survey \citep{Kalberla2005}. This
new survey combines the southern sky survey of the Instituto Argentino
de Radioastronom\`ia (IAR) \citep{Bajaja2005} with an improved version
of the Leiden/Dwingeloo Survey (LDS) \citep{Atlas1997}. Currently this
is the most sensitive Milky Way \hi line survey with the most extensive
coverage both spatially and kinematically. Naturally, since \hi provides
the main emission line of HVCs, one can hope that this survey may
provide some new insights to the HVC phenomenon. A comparison to
previous results, in particular to those by \citet{WvW}, appears
mandatory.

The LDS survey, covering declinations $\delta \ga -30\deg $, is
available since 1997. Its velocity coverage of $ -450 < v < 400$ \kms
is large enough to study most of the high velocity gas in the Milky
Way, only complex EN is known to have \hi emission at $ v < -450$ \kms. 
For HVC research the LDS  was predominantly used for large
scale investigations, e.g. searches for signs of interaction between
HVCs and disk gas \citep{Pietz1996,Bruens2000}, a search for soft X-ray
emission associated with prominent high-velocity-cloud complexes,
\citep{Kerp1999} and a proposal to explain HVCs as building blocks of the
Local Group \citep{Blitz1999}. A new class of compact isolated HVCs
has been detected by \citet{Braun1999} and studied in more detail by
\citet{deHeij2002b,deHeij2002c}. An advantage of the LDS over older
surveys was the improved velocity resolution, also the better  
spatial coverage without gaps. Most of the instrumental errors originating
from the side-lobes of the antenna diagram have been removed. Still,
some instrumental problems are noticeable at a level $T_B \la 0.1$K.
\citet{Wakker2004} claims in addition sensitivity limitations and
discrepancies when comparing maps from the LDS with those derived from
the \citet{HW} database. 

The LAB survey improves on the LDS, residual limitations have been
discussed by \citet{Kalberla2005} and are not expected to be severe
enough to cause notable problems. Still, maps of HVCs based on the LAB
survey \citep[][see e.g. Figs. 4\&5]{Kalberla2005} do not differ
significantly from LDS maps and the discrepancies discovered by
\citet{Wakker2004} remain to be explained. 

In Sect. 2 we compare column densities derived from the LAB survey with
those from \citet{WvW}. We find systematical deviations in line widths
listed by \citet{WvW} and those derived from the LAB as second moments
of the HVC line emission. To study the discrepancies in more detail we
use in Sect. 3 a Gaussian decomposition.  We find for most of the HVC
complexes clear indications for a multi-phase structure and discuss the
properties of individual HVC complexes in Sect. 4 in some detail.  There
are hints for a possible interaction between HVCs and gas closer to the
Milky Way, described in Sect. 5. In Sect. 6 we discuss properties of the
multi-phase HVC gas. Our summary is in Sect. 7.

\section{Basic statistical investigations}

We use measurements and cloud assignments based on the
observation by \citet{HW} and \citet{Bajaja1985}. The database in form
of a table containing HVC positions, velocities, line widths, and peak
temperatures, was prepared by \citet{Wakker1990} and used by
\citet{WvW} for a classification of HVCs into complexes and
populations. Also the review by \citet{WvW97} is based on this table.
We therefore use the term ``WvW database'' or ``WvW table'' for
reference.

\subsection{Peak temperatures and calibration}

For each entry in the WvW database we compare first the peak temperature
at a given center velocity $v_c$ with the temperature of the LAB profile
at the same velocity.  Unfortunately, due to differences in sampling,
only a small number of LAB profiles agree in position exactly with those
entering the more coarsely sampled material used by \citet{WvW}. To
overcome this discrepancy, a mean LAB profile was derived by convolving
the original observed LAB database with a Gaussian smoothing kernel of
0.3\deg~ FWHM. This degrades the angular resolution for the LAB data
slightly (10\%). Next we took the difference in velocity resolution into
account. The spectra available for the WvW catalog had 8 \kms~ velocity
resolution, and were smoothed to 16 \kms~ before their analysis, while
the LAB survey was observed with a resolution of 1.03 \kms. A direct
comparison of the peak temperatures would be affected by resolution and
instrumental noise, we therefore calculate mean temperatures from the
LAB spectra by averaging 15 channels, corresponding to the velocity
resolution of 16 \kms~ of the spectra used by WvW. Comparing the derived
peak temperatures leads to an excellent agreement. On average, the peak
temperatures deviate by 0.9\% only, indicating a perfect consistent
calibration.

\subsection{Direct comparison of column densities}

Next we compare the derived column densities. From the interpolated LAB
profile we determined the \hi column density for each HVC component in
the WvW list by integrating the LAB profile within the velocity range
$v_c - \Delta v < v < v_c + \Delta v$, where $v_c$ and $\Delta v$ are
the LSR component velocity and its associated FWHM velocity width
according to the WvW table.  For a Gaussian distribution 98\% of the
line integral is expected in this range.

\begin{figure}[!ht]
   \centering
   \includegraphics[width=9cm]{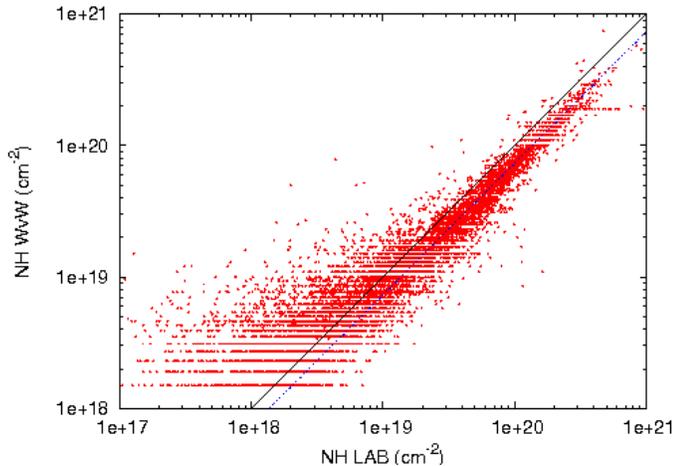}
\caption{\hi column densities from the WvW database in comparison with
column densities determined from the LAB survey. The solid line
indicates the expected correlation, the lower dotted line (blue)
$N_H$(LAB) = 1.37 $N_H$(WvW). }
         \label{FigNH1}
   \end{figure}

Fig. \ref{FigNH1} shows a comparison between column densities, derived
from the LAB, with those from the WvW catalog. We find the regression
$N_H({\rm LAB}) = 1.37 \cdot N_H({\rm WvW})$ (lower dotted line). Despite a
large scatter, there is at large column densities a clear deviation
from the expected relation (solid line). At low column densities the
scatter gets worse, apparently indicating on average $N_H({\rm LAB}) <
N_H({\rm WvW})$. The latter effect might be caused by increasing
uncertainties for component parameters at low column densities,
leading to a systematic mismatch for the redetermined LAB parameters.

\begin{figure}[!ht]
   \centering
   \includegraphics[width=9cm]{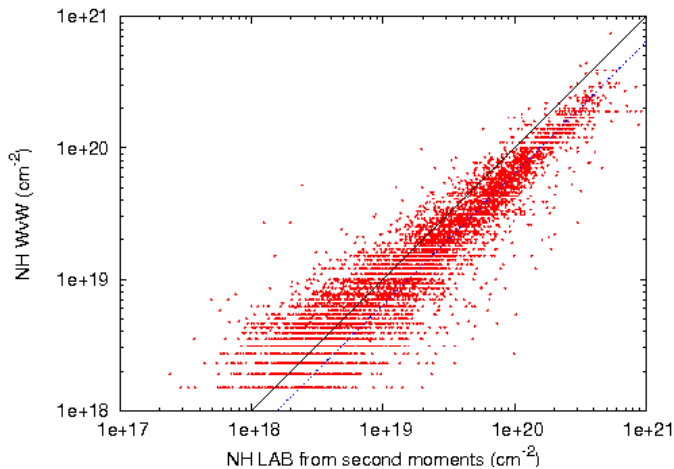}
\caption{\hi column densities from the WvW database in comparison with
column densities determined from the LAB survey after a redefinition of
the HVC component widths. In this case the LAB survey was smoothed to a
resolution of $1.5\deg$. The solid line indicates the expected
correlation, the lower dotted line (blue) the correlation $N_H$(LAB) =
1.55 $N_H$(WvW). }
         \label{FigNH2}
   \end{figure}

\subsection{Column densities from first and second moments}

To investigate the problem in more detail, we refined the
determination of component center velocities and dispersions for the
LAB profiles. We use peak velocities $v_p$, redetermined from the
LAB, and line widths $\Delta v$ from the WvW table. Next we derive for
the range $v_p - \Delta v < v < v_p + \Delta v$ a first guess for the
first and second moments of each HVC component.  Next we replace
$v_p$ with the first moment and $\Delta v$ with the FWHM derived from
the second moment and reiterate the last step to obtain the final
component velocity $v_{cL}$ and its FWHM width $\Delta v_{cL}$. We
then use the velocity range $v_{cL}-\Delta v_{cL} < v < v_{cL}+\Delta
v_{cL}$ for a new determination of LAB column densities. For 85\% of
all components in the WvW list we find this way components with well
defined peak temperatures, first, and second moments. To improve the
signal-to-noise ratio of the LAB survey we repeated the whole
procedure for different spatial resolutions of the LAB profiles.  We
used alternative Gaussian smoothing kernels with FWHM of 1.0\deg~ and
1.5\deg~ and derive $N_H({\rm LAB}) \sim 1.55~ N_H({\rm WvW})$.
Fig. \ref{FigNH2} displays WvW and LAB column densities for a 1.5\deg~
smoothing kernel.

Improving our search algorithm for HVCs with respect to the procedure
described in Sect. 2.2 leads to a significant decrease of the scatter
at low column densities. Comparing the different resolutions we find
that the total scatter, however, increases with increasing LAB beam
width. Spatial fluctuations in the HVC emission may be responsible for
this effect. An increasing mismatch of the beam shapes would cause
increasing discrepancies of the derived column densities.

When comparing the derived center velocities $v_{cL}$ with velocities
$v_c$ from the WvW catalog, we find RMS deviations between 7.7 \kms~
(1.5\deg~ smoothing kernel) and 9 \kms~ (0.3\deg~ kernel). This is
about half of the FWHM velocity resolution and twice the expected
uncertainties of 4 \kms~ \citep{HW}. The deviations are purely
statistical and without any bias. The average component velocities
agree within 0.01 \kms. Spectrometer problems as a reason for the
column density discrepancies can therefore be excluded.  

A major restriction of our analysis is that our component search
algorithm diverges for 15\% of the WvW components. Discrepant
component parameters might have two reasons. The signal in the LAB
survey may be too weak. This applies to 3\% of all cases. In most
cases, 12\%, we found a profile structure that is too complicated for
our approach.  We demand in any case $|v_{cL}-v_c| < \Delta v$, the
refined component velocity has to be within the limits of the WvW
catalog. For a line width of 20 \kms~ this means a 2.5 sigma deviation, a
rather stringent limit.
 
Among those LAB profiles that fail to have HVC emission at a four to
five sigma level at positions where WvW components are present we
visually inspected 115 of the most serious cases. In 29 cases we find
some emission at velocities deviating significantly from the WvW catalog
but in 74 cases we do not detect any significant HVC emission in the LAB
profiles. For the remaining 12 cases the situation was unclear, the LAB
baselines might have been affected by interference.

\subsection{Comparison of line widths}

\begin{figure}[!ht]
   \centering
   \includegraphics[width=9cm]{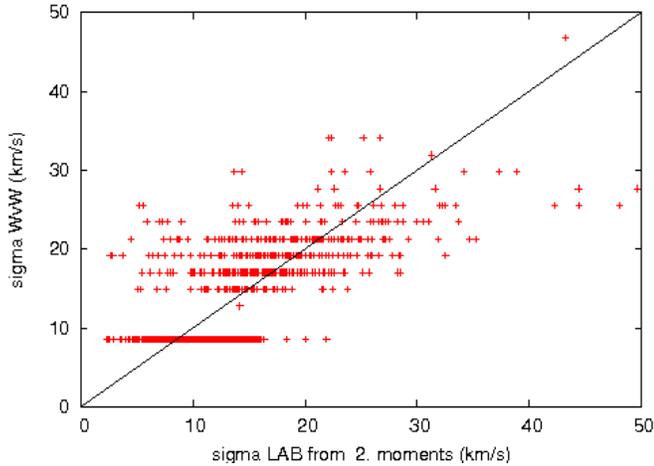}
\caption{Estimated velocity dispersions from the WvW database in
comparison with dispersions determined from the LAB survey using
second moments. In 87\% of the WvW database the estimate $\sigma = 8.5$ \kms~
(FWHM = 20 \kms) was used, too crowded to be resolved in this plot. }
         \label{Figmom}
   \end{figure}

Searching for a reason for the discrepancies in the derived column
densities we can exclude calibration and spectrometer problems, but a
discussion of the linewidth remains to be done.  Fig. \ref{Figmom}
displays a comparison of velocity dispersions, which implicitly have
been used to calculate the column densities in Fig. \ref{FigNH2}. To
derive this Fig. we degraded the spatial resolution of the LAB to
1.5\deg. Our intention was to improve the sensitivity of our analysis
without degrading the velocity resolution. In turn, uncertainties for
the derived second moments of the HVC components depend on their
widths. Components with dispersions $\sigma \ga 8.5$ \kms~ are in
sensitivity equivalent to the components from the \citet{WvW}
database. This, however, applies to more than 90\% of all cases, the
huge scatter in Fig.  \ref{Figmom} is significant. In particular, we
note that 87\% of all WvW components have a dispersion of 8.5
\kms. Most of our second moments are significantly larger. Fig.
\ref{Figmom} shows this effect barely because of the crowding of the
WvW components with a dispersion of 8.5 \kms.

Line widths in the WvW database were based on observations with a poor
velocity resolution. The spectra were hanning smoothed to 16.5 \kms
resolution and the data quality did not allow to distinguish lines of
25 \kms FWHM or narrower. As estimates were needed (Wakker, private
communication), these were based mostly on the widths of the profiles
in cloud envelopes that were found in several 1970s papers, such as
\citet{Cram1976,Giovanelli1976,Giovanelli1977,Davies1976}.  Originally
\citet{HW} have chosen a line width of 25 \kms FWHM but \citet{WvW}
revised this to 20 \kms, corresponding to a dispersion of $\sigma =
8.5$ \kms, the value we use here. 

We conclude that the scatter in
Fig. \ref{Figmom}, independent from the signal-to-noise level of the
HVC component, originates predominantly from uncertainties in
estimating the line width. The bias $N_H$(LAB) $\sim 1.5 N_H$(WvW),
visible in Fig.  \ref{FigNH1} \& \ref{FigNH2} would be explainable as
a general bias of the the narrowest lines which are most frequent in
the WvW table. 

We verified, whether non-Gaussian line shapes might affect the
linewidth determination. Using channels centered on the component
velocity $v_c$ with a window given by the FWHM $\Delta v$ on both
sides of the line should, in case of a Gaussian component, lead to
98\% of the line integral. Broadening the window should essentially be
without significant effects. We tested this assumption, increasing the
window by 10\%. As a result the line integrals increase 2.7\%, but
only 1\% is expected. We conclude that HVC components, on average,
must have line shapes that deviate significantly from single component
Gaussians.  As an additional test we have split the HVC components at
their center velocities in two parts. Both subcomponents were
integrated separately. In this case the discrepancies increased by
another factor of two, indicating that a significant fraction of the
HVC line components must be asymmetric. From high resolution
observations it is known since long that HVC lines are asymmetric
(e.g. \citet{Giovanelli1973}, see also the discussion in Sect. 2.7 of
\citet{WvW97}). Apparently we recovered this general property of HVCs
on a broader basis. Some uncertainties are expected if column
densities are calculated by simply multiplying the peak of the line
emission with a measure of the component width.

\section{Gaussian decomposition} 

We decomposed all profiles of the LAB survey into Gaussian components.
The procedure is equivalent to the decomposition of the LDS described by
\citet{Haud2000}. A detailed statistical analysis and a
classification of Gaussian components concerning possible spurious
components is given by \citet{Haud2006}. A publication of the database
is in preparation. Based on the WvW database we searched for
components within the LSR velocity range $v_c - \Delta v < v < v_c +
\Delta v$. All positions within the HVC complexes have been selected by
interpolating the WvW HVC distribution. According to the
predefined velocity range we rejected components with $ |v| < 90$ \kms,
exceptions will be discussed in Sects. 4 to 6. Also components with
$\sigma > 50$ \kms~ or $\sigma < 0.5$ \kms~ have been rejected because
they most probably represent spurious signals due to baseline
deficiencies or single-channel interference spikes.

\subsection{Line width distribution}

\begin{figure}[!ht]
   \centering
   \includegraphics[width=9cm]{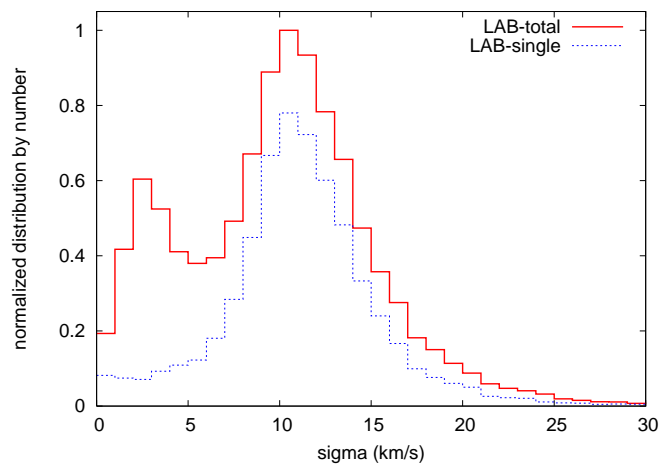}
\caption{Frequency distribution of component line widths. The solid line
is the histogram derived after Gaussian decomposition of the LAB
survey. The lower dotted line represents the subset of single component
HVC profiles, using the same normalization.}
         \label{FigGaussn}
   \end{figure}

The frequency distribution of derived LAB Gaussian components with
respect to their velocity dispersions is shown in
Fig. \ref{FigGaussn}. The upper distribution shows a summary over
all 23070 components found at 17336 positions while the lower dotted
distribution is restricted to those 13221 positions where only a single
Gaussian component was found. Clearly, broad lines dominate the
distribution, in particular for single component features. The first
moment of the upper distribution is at 10.7 \kms. Narrow components with
typical dispersions of 3 \kms~ are most significant at positions where
a multi-phase structure is found.

\begin{figure}[!ht]
   \centering
   \includegraphics[width=9cm]{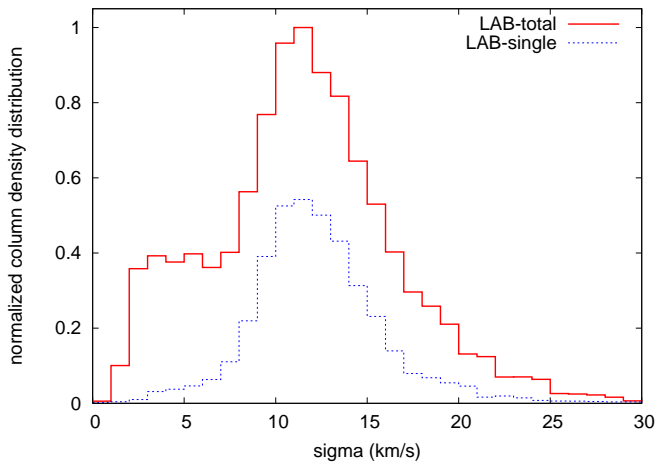}
\caption{Distribution of column densities as function of linewidth.  The
solid line is the histogram derived from the LAB survey. The lower
dotted line represents the subset of single component HVC profiles,
using the same normalization.}
         \label{FigGaussNH}
   \end{figure}

Fig. \ref{FigGaussNH} shows how the Gaussian components contribute to
the observed column densities. At positions having only a single
component (lower dotted line) we barely find narrow lines.  62\% of the
total column density is associated with multi-component structures, but
these cases are observed only at 24\% of all positions.  Broad lines
dominate also the multi-component distribution, 75\% of the
multi-component HVC emission is in broad lines.  We call therefore the
broadest line of a multi-component set the ``primary'' component. This
definition allows us to differentiate Gaussian components according
their line widths. The narrow lines are the ``secondary'' components.
If several secondary components are found we sort according to their
velocities. Components with the least deviations in velocity are
considered first.

The secondary components contain in most cases less column density than
the primary ones.  Nevertheless, secondary components are
important. They contain only 21\% of the total observed \hi column
density but are tracers for regions with significant enhanced HVC
emission, resembling ``\hi icebergs''. The first moment of the total observed
distribution (upper curve in Fig. \ref{FigGaussNH}) is $\sigma = 11.8$
\kms. For those spectra that have single HVC components only (lower
curve in Fig. \ref{FigGaussNH}) we find a first moment $\sigma = 11.1$
\kms.

First evidence for a multi-component velocity structure of some HVCs has
been given by \citet{Giovanelli1973}, a more detailed analysis was
provided by \citet{Cram1976}. These authors used the NRAO 140-foot
telescope to search about 100 directions with local concentrations of
HVC emission for multi-component features.  They found evidence for two
well defined velocity domains, one at FWHM of about 23 \kms~ ($\sigma
\sim 10~ $ \kms) and one at FWHM of about 7 \kms~ ($\sigma \sim
3 $ \kms). Broad components, called by us ``primary'', correlate
with more extended regions (envelopes). Narrow lines, ``secondary''
components, correlate with small, bright condensations (cores) and with
larger column densities. This finding resembles the well established
two-component structure of the neutral interstellar medium. 

Our results are in excellent agreement with \citet{Cram1976}, except
that we find somewhat larger line widths. A reason may be that
\citet{Cram1976} disregarded components associated with line wings,
excluding broad lines with $\sigma > 13$ \kms~ from their analysis.
\citet{Lockman2002}, pointing the NRAO 140-foot telescope to 860 random
positions, found a distribution very similar to our single component
distribution, with median dispersion of 12.9 \kms. Narrow lines may be
underrepresented in their analysis due to limitations in velocity
resolution of 5 \kms~ FWHM. \citet{Burton2001} derived a typical width
of $\sigma \sim 10.6$ \kms~ from Arecibo observations of compact high
velocity clouds. There might be a tendency that the component widths
decrease with decreasing beam width but $11 \la \sigma \la 13$ \kms~
appears to be a typical dispersion, our result is in any case
consistent.

87\% of the WvW database has $\sigma = 8.5$ \kms~ (FWHM = 20 \kms),
quite different from our results. Comparing the line widths, we find a
typical ratio $\sigma({\rm LAB})/\sigma({\rm WvW}) \sim 1.4$. Recalling
that we found an excellent agreement in the observed peak temperatures,
we explain the systematic \NH~ discrepancies discussed in Sect. 2 (see
also Figs. \ref{FigNH1} and \ref{FigNH2}) as due to a general bias in
the WvW line widths.

\begin{figure}[!ht]
   \centering
   \includegraphics[width=9cm]{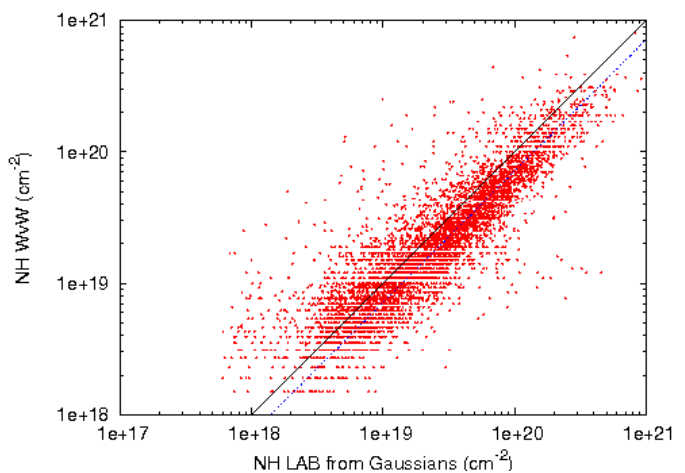}
\caption{\hi column densities from the WvW database in comparison with
column densities determined from the LAB survey after Gaussian
decomposition.  For each position from the WvW table the nearest LAB
position was searched for.  The solid line indicates the expected
correlation, the lower dotted line (blue) the correlation $N_H$(LAB) =
1.40 $N_H$(WvW).  }
         \label{FigNHG1}
   \end{figure}

\begin{figure}[!ht]
   \centering
   \includegraphics[width=9cm]{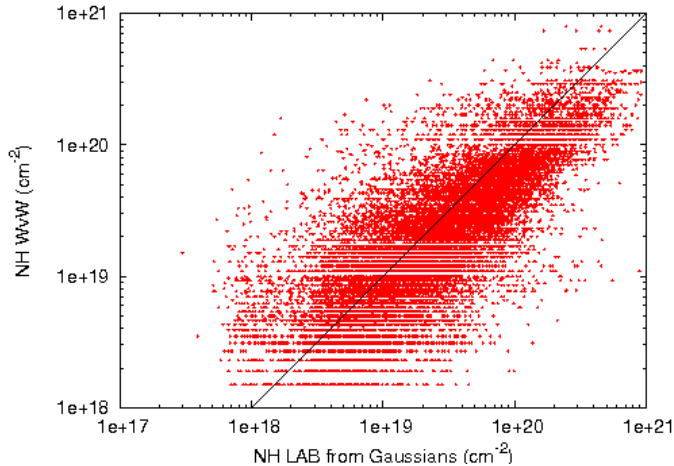}
\caption{\hi column densities from the WvW database in comparison with
column densities determined from the LAB survey after Gaussian
decomposition.  All LAB profiles within HVC complexes according to WvW
have been used.  The solid line indicates the expected correlation.  }
         \label{FigNHG2}
   \end{figure}

\subsection{Column densities from Gaussian components}

We use the Gaussian components and display in Figs.  \ref{FigNHG1} a
comparison of LAB component column densities with those from the WvW
catalog. The linear regression line shows $N_H({\rm LAB})/N_H({\rm WvW})
= 1.4$, consistent with the previous discussion.  For Fig. \ref{FigNHG1}
we use for each WvW position the nearest LAB position only. For
Fig. \ref{FigNHG2} all of the LAB profiles next to each WvW position are
considered, thus filling the 1.0\deg~ \citep{HW} or 2.0\deg~ grid
\citep{Bajaja1985}. If the HVC distribution were smooth, both plots
would be identical. However, the large scatter in Fig. \ref{FigNHG2}
degrades the expected correlation.  A comparison of this kind is
therefore highly problematic. In the same way it is quite a problem to
compare LAB column density maps with maps generated from highly
under-sampled observations on a 1.0\deg~ or 2.0\deg~ grid. This was
noted already by \citet{Wakker2004}. Consistent maps were obtained only
after smoothing the LDS to a 1.0\deg~ beam and using a careful selected
velocity range.

In 1991 the WvW database was the best all sky HVC catalog and formally it
is still the most sensitive one.  The one-sigma noise limit is 10 mK
\citep{HW} for the northern and 25 mK \citep{Bajaja1985} for the
southern part but possible systematic errors have not been taken into
account, profile wings in particular may be affected. For the LAB
survey \citep{Kalberla2005} the side-lobe response was removed and
residual errors are most probably below a level of 20 -- 40 mK, an
order of magnitude below typical errors for radio telescopes.  Some
compromises had to be made by \citet{WvW}, notably on the velocity
resolution and the sampling interval. Uncertainties in center
velocities and velocity dispersions cause problems when selecting a
velocity range. Moreover, the LAB survey shows significant {\it real}
fluctuations. Under-sampling necessarily must cause
inconsistencies. \citet{HW} estimate that their catalog is 57\%
complete for clouds with a central peak brightness of 0.2 K. Smoothing
the LAB to 1.0\deg~ resolution helps a little when comparing with
previous HVC maps \citep{Wakker2004} but inconsistencies caused by an
invalid interpolation of missing data in the WvW database due to a
grid of 1\deg~ or 2\deg~ cannot be overcome.

\section{Multi-component statistics}

The previous sections have shown that the LAB survey has major
advantages in its completeness on scales of 0.5\deg and in particular
in its high velocity resolution, allowing an unbiased analysis for
most of the observable HVC emission. Using a Gaussian decomposition
\citep{Haud2000}, we found a large number of positions with a
multi-component structure.  In the following we shall focus on the
properties of these multi-component features. We distinguish complexes
according to the most recent classification by \citet{Wakker2004} and
find substantially different structures for individual complexes. This
makes a discussion of individual complexes necessary. To keep this
contribution comprehensive we shall focus on complex A, most of the
other complexes are mentioned only briefly. On first reading
individual subsections may be skipped.

\subsection{Complex A}

Complex A is one of the first discovered HVCs
\citep{Muller,Hulsbosch66}. It's distance is 8 to 10 kpc, the metallicity
probably close to 0.1 times solar
\citep{vanWoerden1999,Wakker2001,vanWoerden2004}. It's velocity structure
is complex, condensations tend to be associated with higher velocities 
\citep[see e.g. Fig. 4 of ][]{Wakker2001}. 

\begin{figure}[!ht]
   \centering
   \includegraphics[width=9cm]{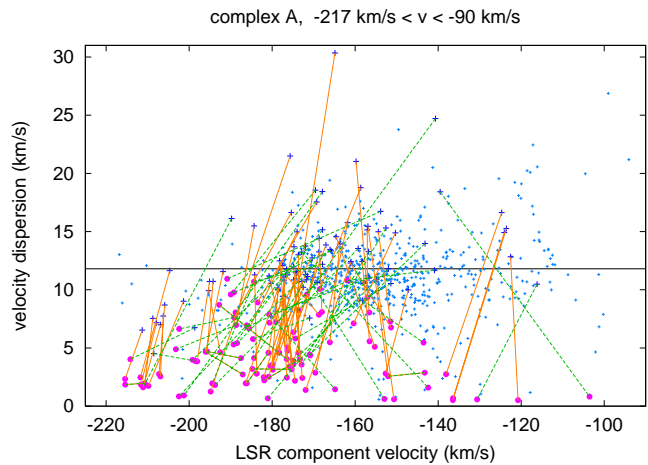}
\caption{Distribution of the velocity dispersions of individual Gaussian
components with respect to the observed LSR component velocities.  The
horizontal line indicates the mean dispersion $\sigma = 11.8$
\kms. Crosses (blue) indicate primary, dots (magenta)
secondary components. Solid lines (orange) connect components with
$|v_i-v_1| < \sigma_1$, dotted lines (green) with $|v_i-v_1| < {\rm
FWHM}_1$. }
\label{FigA_vau-sig}
   \end{figure}

Fig. \ref{FigA_vau-sig} shows the distribution of component velocity
dispersions derived from the LAB survey with respect to the observed
center velocities. The crosses indicate primary, the dots secondary
components. At high velocities, $ v \la -180$ \kms, we find
predominantly narrow lines with dispersions lower than the average
dispersion $\sigma = 11.8$ \kms.  For $ v \ga -150$ \kms predominantly
high velocity dispersions show up.  To display the relations between
primary and secondary components, observed at the same position, we have
drawn a connection between these components. Solid lines were used in
cases when the center velocity of the secondary component (index i)
deviates from the primary component (index 1) by less than its
dispersion, $|v_i-v_1| < \sigma_1$. For the hypothesis of a core moving
within an envelope this indicates subsonic motions.  Supersonic motions
with $|v_i-v_1| < 2.35 \cdot \sigma_1$, corresponding to the FWHM line
widths of the primary component, are indicated by dotted lines. We
disregard larger Mach numbers $M = |v_i-v_1| / \sigma_1$, as
Fig. \ref{Figa_histo_S} shows that these are unimportant.

\begin{figure}[!ht]
   \centering
   \includegraphics[width=9cm]{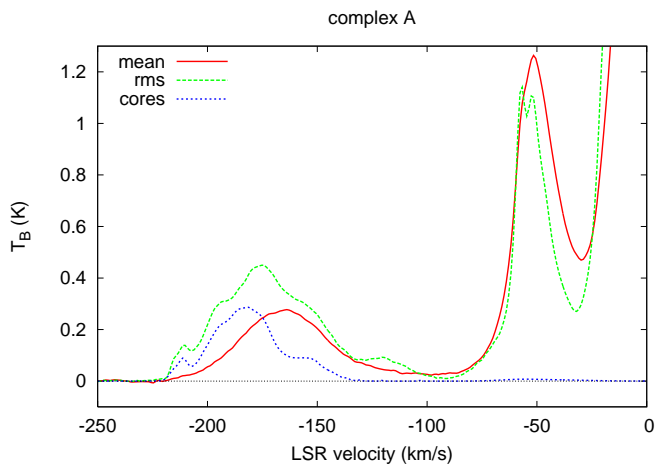}
\caption{Mean emission (solid line, red) and rms fluctuations (dashed
line, green) from the LAB survey calculated for all positions where HVC
emission was detected.  The lower dotted line (blue) shows the mean
emission from the secondary Gaussian components. }
\label{FigA_rms}
   \end{figure}

To demonstrate that secondary components unambiguously are related
with small scale features we compiled for complex A three spectra, the
mean \hi emission of this complex for all LAB positions where HVC
emission was detected, the corresponding rms deviation from the mean,
and in addition the mean emission of the secondary components. The rms
spectrum is most sensitive for small scale features, emission on
scales $ d \ga 0\fdg5$ is filtered out \citep[for discussion of the
properties of rms spectra see e.g.][]{Mebold1982}. Fig. \ref{FigA_rms}
shows that the emission from secondary HVC components is similar to
the rms spectrum, except for an arbitrary scaling depending on the
telescope beam width. The situation is different at intermediate
velocities. Secondary IVC components, possibly associated with primary
HVC components, are negligible.

\begin{figure}[!ht]
   \centering
   \includegraphics[width=9cm]{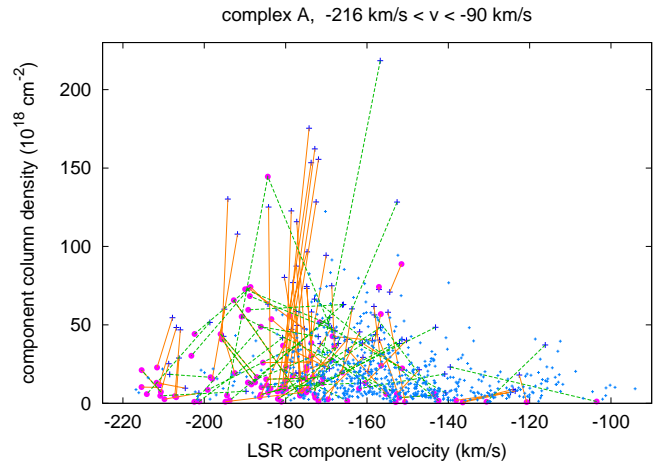}
\caption{Distribution of component column densities 
with respect to the observed LSR component velocities.
Crosses (blue) indicate primary, dots (magenta) secondary components. 
Solid lines (orange) connect components with $|v_i-v_1| <
\sigma_1$, dotted lines (green) with $|v_i-v_1| < {\rm FWHM}_1$.}
\label{FigA_vau-NH}
   \end{figure}

\begin{figure}[!ht]
   \centering
   \includegraphics[width=9cm]{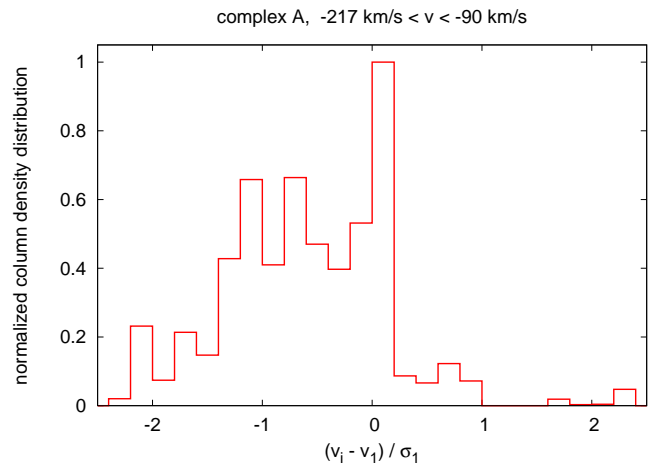}
\caption{Column density weighted frequency distribution of $(v_i -
v_1) / \sigma_1 $. }
\label{Figa_histo_S}
   \end{figure}

\begin{figure}[!ht]
   \centering
   \includegraphics[width=9cm]{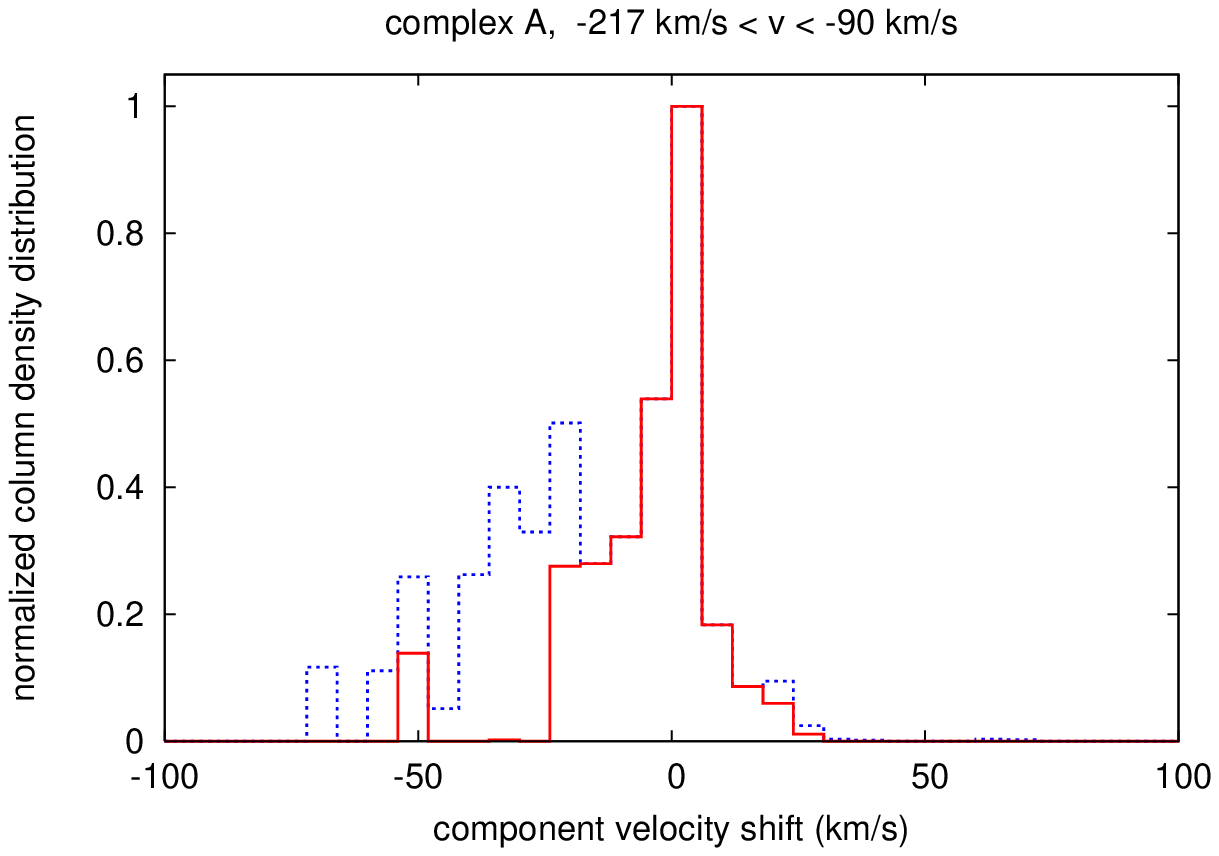}
\caption{Column density distribution of secondary components 
with respect to their velocity shifts $v_i-v_1$. 
Solid lines are for subsonic components with $|v_i-v_1| < \sigma_1$, dotted
lines for supersonic components up to Mach 2.35. }
\label{FigA_histo_V}
   \end{figure}

In Fig. \ref{FigA_vau-NH} we display the HVC component column densities
against their velocities. Symbols and lines have the same meaning as in
Fig. \ref{FigA_vau-sig}. For $ v \la -170$ \kms, a significant fraction
of the HVC emission is contributed by secondary components.  At $ v \ga
-150$ \kms~ secondary components are insignificant but also the primary
components have low column densities. Multi-component structure in
complex A exists only at 13\% of the positions but most of this is at
high velocities.

Figs. \ref{FigA_vau-sig} \& \ref{FigA_vau-NH} show a complicated
dynamical network in 3-D, relating velocities, line widths and column
densities between primary and secondary components. Some substructures
appear to repeat. In particular, we find that many of the lines 
connecting primary and secondary components in Fig. \ref{FigA_vau-sig}
run parallel.

Fig. \ref{Figa_histo_S} shows the frequency distribution of $(v_i
- v_1) / \sigma_1 $. A comparison between Fig. \ref{FigGaussn} and
\ref{FigGaussNH} demonstrates that the statistic depends on the
weight. Rather then using number statistics we decided to weight for
the rest of the paper components according to their column
densities. To allow a better comparison we normalize in general the
peak to 1. We find a bimodal situation. The narrow peak at
$(v_i-v_1)/\sigma_1 \sim 0$ corresponds to cores with random motions
and Mach numbers $ M \la 0.5$.  A broad distribution with $ 0.5 \ga M
\ga 1.5 $ exists for cores at more negative velocities than the
primary components.  In Fig. \ref{FigA_histo_V} we plot the column
density weighted distribution of the component velocity shifts
$v_i-v_1$. The solid line represents components with $|v_i-v_1| <
\sigma_1$, the dotted line is for the supersonic case. This histogram
again shows a clear pattern. The peak around zero is due to the cores
with the low Mach numbers. Another group of cores exist with velocity
shifts up to $-50$ \kms, these are the clumps with larger Mach
numbers.

The first moment of this distribution is at $v_i-v_1 = -3.7 $ \kms~ for
subsonic components and $v_i-v_1 = -7.5 $ \kms~ for supersonic components
(see Table 1).  For a multi-phase medium in equilibrium one would
expect a symmetrical distribution of component column densities around
zero velocity shift.  If a multi-component interpretation is valid for
complex A we conclude that this HVC must be in a highly non-equilibrium
state.

\subsection{Complex C}

Complex C is the second largest HVC complex and is at a
distance $ \ga 6$ kpc. The metallicity is $\sim 0.1 - 0.2 $ solar
\citep{Wakker2004,Tripp2003}. This complex has a broad distribution of \hi gas
between $-140 \la V \la -110$ \kms~ and some elongated filamentary
structures at $v \la -140$ \kms. Multi-component structures are
less prominent in comparison to complex A, see Table 1.

\begin{figure}[!ht]
   \centering
   \includegraphics[width=9cm]{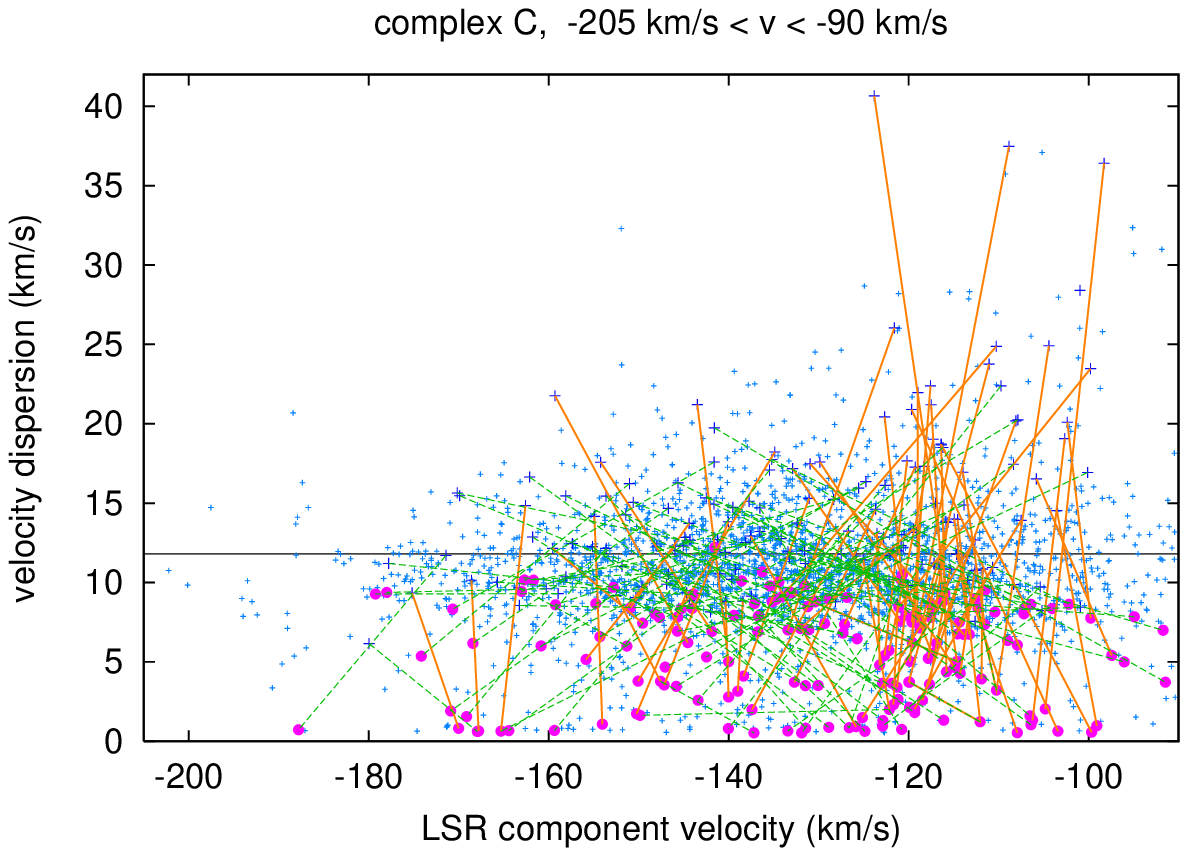}
\caption{Same as Fig. \ref{FigA_vau-sig} but for complex C. }
\label{FigC_vau-sig}
   \end{figure}

\begin{figure}[!ht]
   \centering
   \includegraphics[width=9cm]{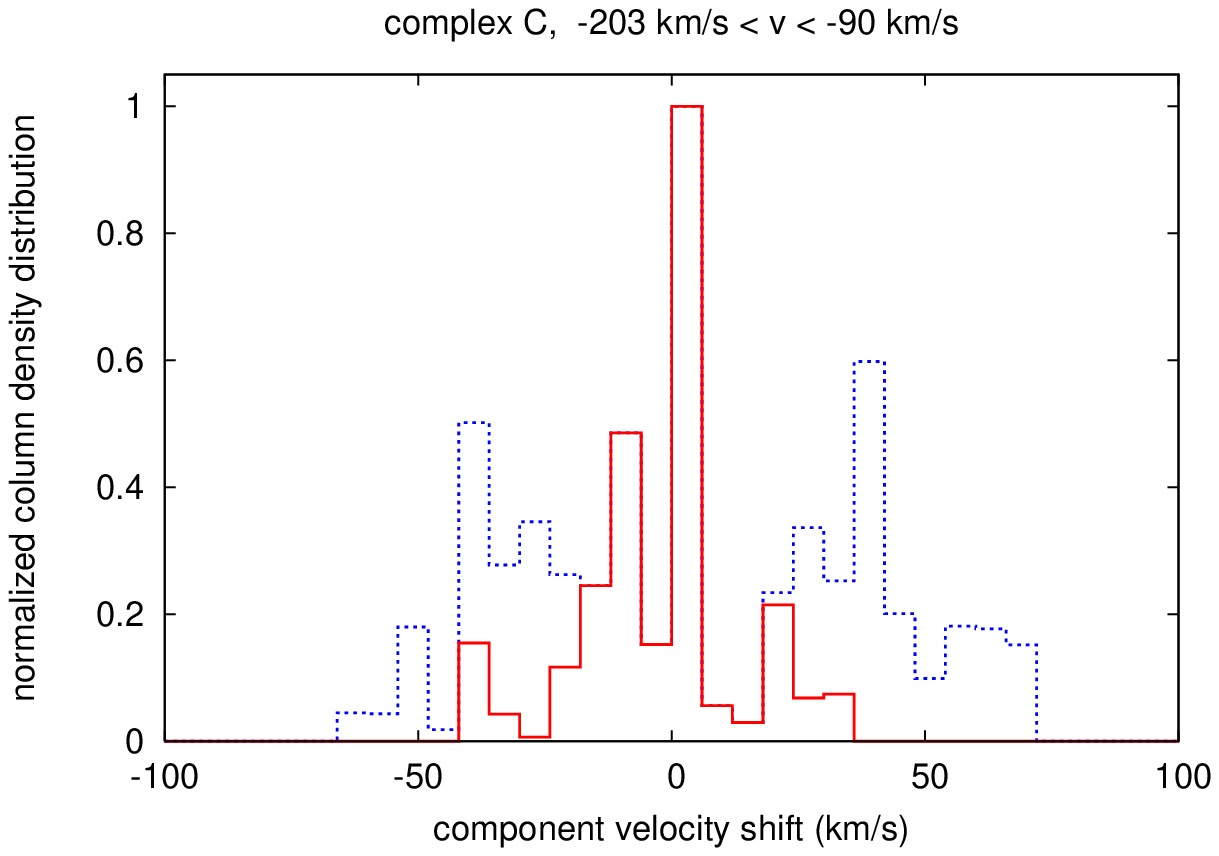}
\caption{Column density distribution of secondary components 
with respect to their velocity shifts $v_i-v_1$. 
Solid lines are for subsonic components with $|v_i-v_1| < \sigma_1$, dotted
lines for supersonic components up to Mach 2.35. }
\label{FigC_histo_V}
   \end{figure}

Fig. \ref{FigC_vau-sig} shows the distribution of velocity dispersions
and the relation between primary and secondary components. These appear
to be distributed randomly, also we find no clear preferences in the
column density distribution, much in contrast to
Figs. \ref{FigA_vau-sig} \& \ref{FigA_vau-NH}.  Velocity shifts between
secondary and primary components are distributed in a nearly symmetric
way, see Fig. \ref{FigC_histo_V}.

\subsection{Complex H}

\begin{figure}[!ht]
   \centering
   \includegraphics[width=9cm]{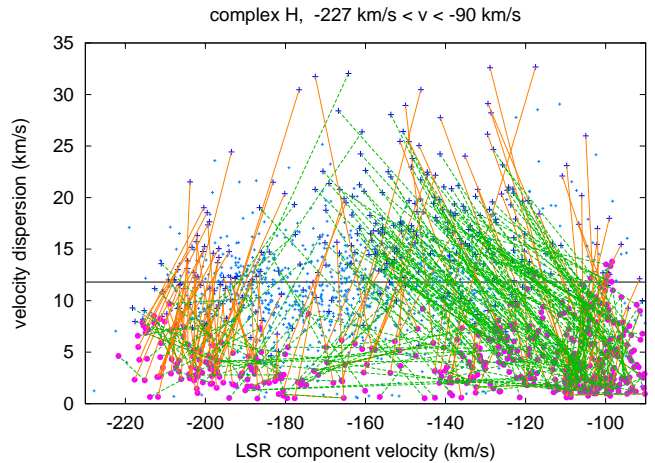}
\caption{Same as Fig. \ref{FigA_vau-sig} but for complex H. }
\label{FigHH_vau-sig}
   \end{figure}

\begin{figure}[!ht]
   \centering
   \includegraphics[width=9cm]{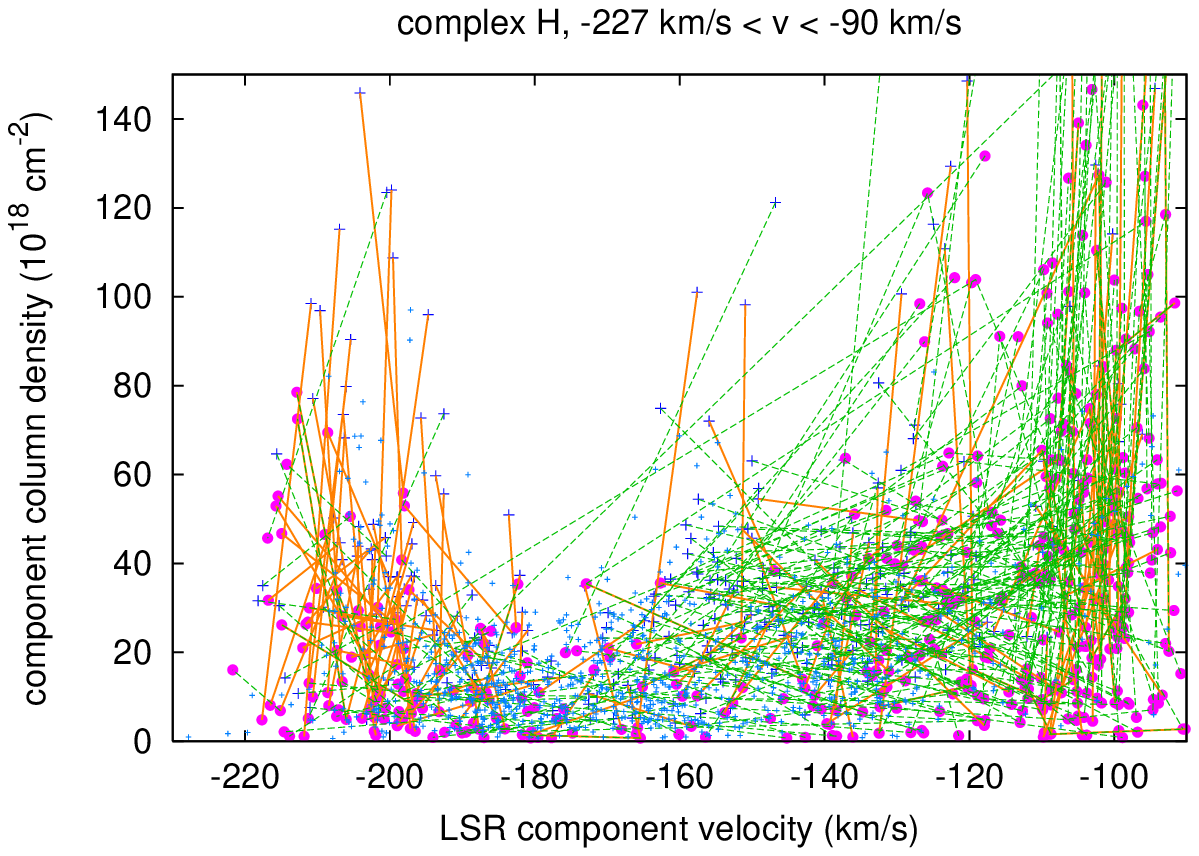}
\caption{Same as Fig. \ref{FigA_vau-NH} but for complex H. }
\label{FigHH_vau-NH}
   \end{figure}

\begin{figure}[!ht]
   \centering
   \includegraphics[width=9cm]{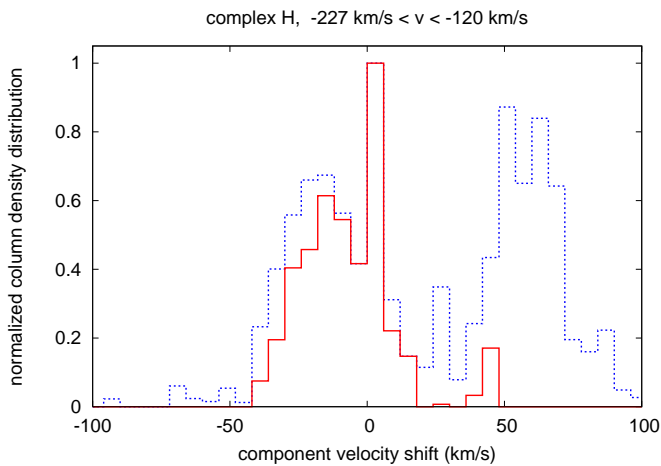}
\caption{Column density distribution of secondary components 
with respect to their velocity shifts $v_i-v_1$. 
Solid lines are for subsonic components with $|v_i-v_1| < \sigma_1$, dotted
lines for supersonic components up to Mach 2.35. }
\label{FigH_histo_V}
   \end{figure}

This complex, studied first by \citet{Hulsbosch1975}, lies in the plane
of the Galactic disk and has an angular extent of $\sim 25\deg$.
\citet{Blitz1999} proposed that it is located at a galacto-centric
distance of $R \ga 40$ kpc. \citet{Lockman2003} advocated 
$R = 33 \pm 9$ kpc. There are some different opinions which part of
the emission should be related to the HVC or the disk. According to
\citet{WvW} emission at $v < -80$ \kms~  belongs to the
HVC. \citet{Wakker2004} considers only gas at $v < -90$
\kms. \citet{Blitz1999} pointed out that at least some of this gas
belongs to the outer edge of the Milky Way disk. 

First we consider the HVC with velocities $v < -90$ \kms.
Fig. \ref{FigHH_vau-sig} shows the distribution of component velocity
dispersions with respect to the observed center velocities.  The
velocities for most of the cores are systematically offset from their
envelopes. Secondary components at $-220 \la v \la -190$ \kms~ are
associated with envelopes at more positive velocities. The opposite
happens for secondary components at $-140 \la v $ \kms, these are
associated with with envelopes at more negative velocities.
Fig. \ref{FigHH_vau-NH} shows that cores with the most extreme
velocities have the largest column densities. Cores at $-120 \le v $
\kms~ have exceedingly large column densities, cut off in the plot. This
part of the emission is most probably caused by the disk, for discussion
see Sect. 6. We therefore 
exclude components at  $v \le -120 $ \kms~ for the further discussion, 
the entry in Table 1 is H(--120). 

Most of the components with $v \la -190$ \kms~ are subsonic. Plotting
in Fig. \ref{FigH_histo_V} the component column density distribution
for $v < -120$ \kms~ with respect to velocity shifts $v_i - v_1$ we
find a very broad distribution. The components group at several
distinct velocity shifts $v_i - v_1$ , most prominent subsonically at
$v_i - v_1 \sim -20 $ \kms~ and supersonically at $v_i - v_1 \sim 60
$ \kms. These are mostly cores with $v \ga -150$ \kms.

Our interpretation of two components at different velocities in
the same direction as physically related gas having subsonic or
supersonic relative motions is not unique. As pointed out by the
referee, the possibility exists that these components are well
separated in distance. The bimodal nature of the shift suggests that
we are looking at two objects, or at least at one object that is
stretched and convoluted such that the sight-line crosses it twice.

Such an interpretation demands a spatial separation of primary and
secondary components. A strict separation would be in conflict with the
general finding that narrow lines are always associated with broad
components, leading to the core-envelope hypothesis
\citep{Giovanelli1973}. Cold components in the halo cannot exist on
their own \citep{Wolfire1995b}. The assumption of an object which has
two or more separate components along the line of sight demands that we
need to take at least two broad components into account. These lines
could blend to a single primary component but we expect in this case
larger line-widths. Table 1 lists in column 9 the first moment of the
column density weighted distribution of velocity dispersions. The entry
for complex H, $\sigma_{NH} = 12.0 $ \kms~ does not deviate
significantly from the average $ <\sigma_{NH}> = 11.8 $ \kms~ for all
Gaussians. Considering only those positions in complex H that have no
associated secondary components, we obtain $\sigma_{NH} = 11.0 $ \kms,
again identical with the value obtained for the whole sample. There are
no indications that our interpretation is affected by a complicated
spatial substructure. We generalize this result for most of the
complexes. MS, C, WA, and WC may be affected for some of the
sight-lines. The interface region close to the Magellanic System most
probably has a complicated spatial structure, we consider this in
Sect. 4.5.

\subsection{Complex M}

\begin{figure}[!ht]
   \centering
   \includegraphics[width=9cm]{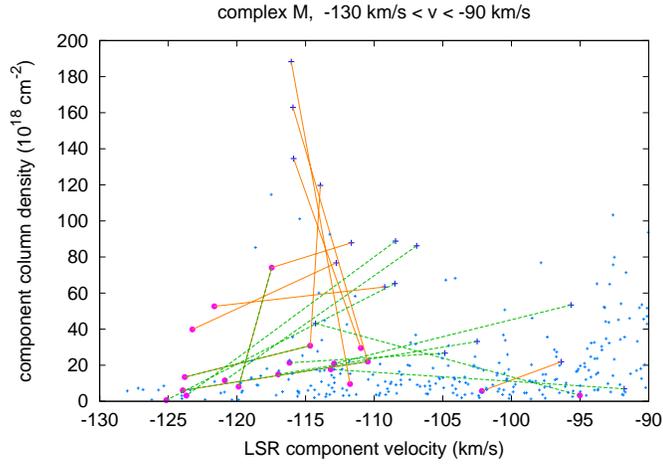}
\caption{Same as Fig. \ref{FigA_vau-NH} but for complex M. }
\label{FigM_vau-NH}
   \end{figure}

\citet{Danly1993} determined a distance to complex M of $ 1.5 \la z \la
4.4$ kpc.  \citet{Wakker2001} cites a metallicity of $\sim0.8$ solar.
In complex M the average velocity dispersion of the Gaussian components
is low, $\sigma \sim 10.3$ \kms. We find only 19 positions with
multi-component structure. The secondary components tend to be
associated with large column densities (Fig. \ref{FigM_vau-NH}).  The
velocities of most of the cores are shifted to larger velocities.

\subsection{Complex MS}

\begin{figure}[!ht]
   \centering
   \includegraphics[width=9cm]{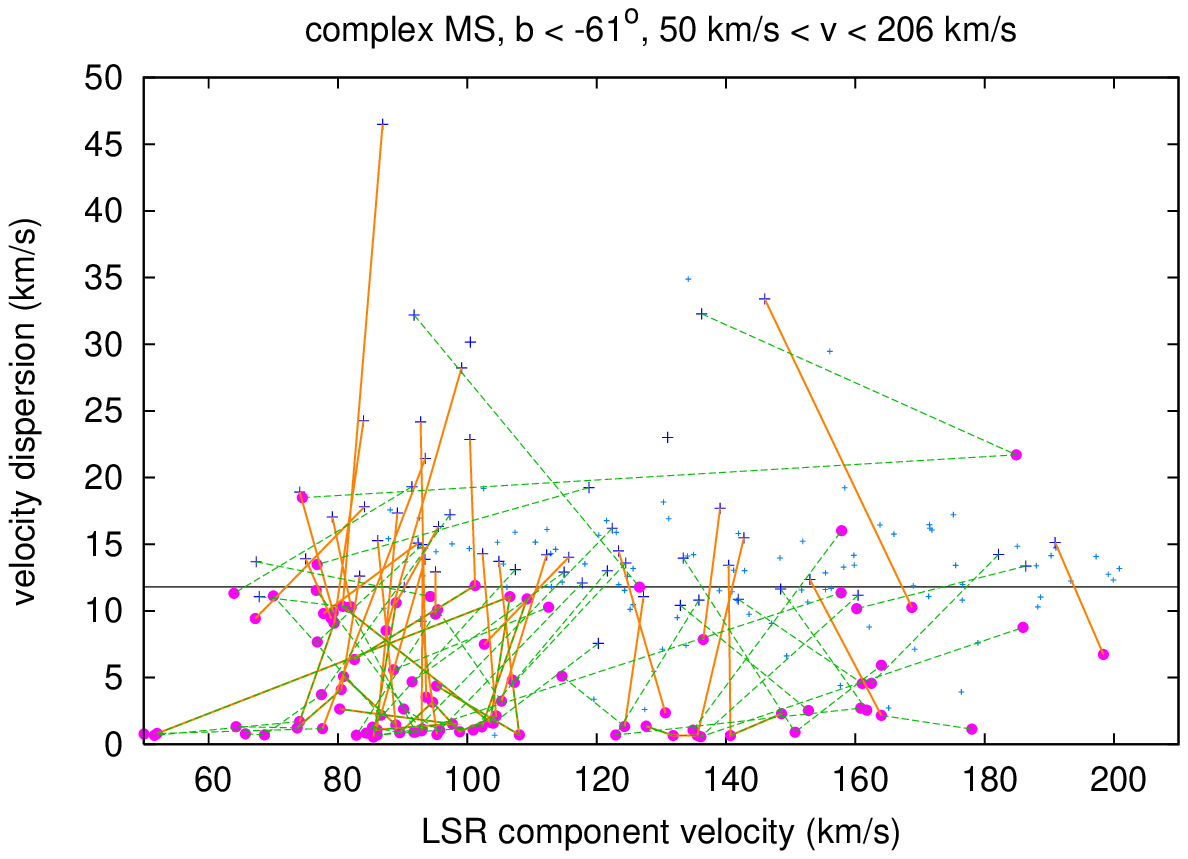}
\caption{Same as Fig. \ref{FigA_vau-sig} but for complex M. }
\label{FigMS+_vau-sig}
   \end{figure}

\begin{figure}[!ht]
   \centering
   \includegraphics[width=9cm]{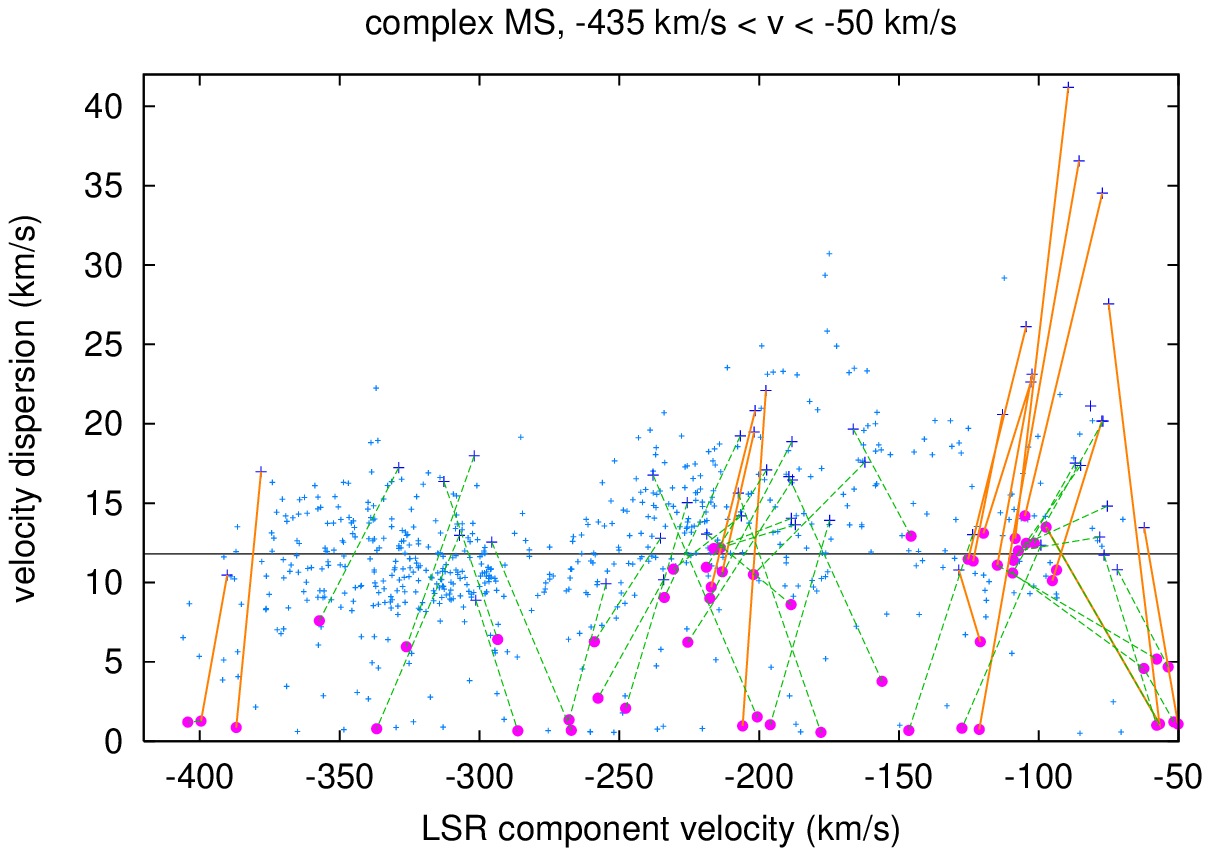}
\caption{Same as Fig. \ref{FigA_vau-sig} but for complex M.  }
\label{FigMS-_vau-sig}
   \end{figure}

The Magellanic Stream can be explained as a tidal tail of the Small
Magellanic Cloud \citep{Gardiner1996}. According to \citep{Wakker2004}
it extends over 120\deg~ and the LSR velocities range from $-440$ to $+412$
\kms. Following \citet[][Figs. 3,4]{Bruens2005} we restricted this range
significantly. We excluded any emission possibly related to the LMC or
the SMC, also to the interface region by selecting a latitude limit $b <
-61\deg$. In addition we splitted this huge complex into two parts with
positive and negative LSR velocities.  In both cases we deviate from the
condition that Gaussian components should have $ |v| > 90$ \kms. We
allow $ |v| > 50$ \kms since we found no problems caused by confusion
with the local gas.  The positive velocity part, MS+, contains 27\% of
the gas in secondary components (Fig. \ref{FigMS+_vau-sig}), while the
negative velocity part, MS--, has only 10\% of the gas in cores
(Fig. \ref{FigMS-_vau-sig}). MS+ shows some scatter in the velocity
shifts but no trend on average, some of the clumps have large column
densities. The properties of secondary components in MS-- are quite
different.  The velocities of the secondary components are significantly
higher, on average shifted by $-9$ \kms~ with respect to the primaries. 

In Table 1 we included for comparison the entry MS+W, according to the
definition by \citet{Wakker2004}, unrestricted with respect to the
interface region. Also in this case a significant fraction of the HVC
emission is found in secondary components. Excluding the interface
region has cut off 20\deg~ of the stream and restricted the velocities to
$v \la 206$ \kms~ but did not affect the statistical properties of its
Gaussian components significantly.

\subsection{Complex EP}

\begin{figure}[!ht]
   \centering
   \includegraphics[width=9cm]{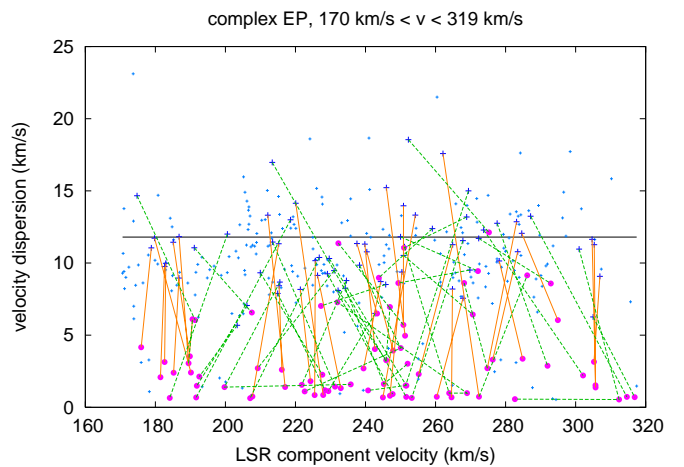}
\caption{Same as Fig. \ref{FigA_vau-sig} but for complex EP. }
\label{FigEP_vau-sig}
   \end{figure}

This is the leading arm of the Magellanic Stream. According to
\citet{Wakker2004} the LSR velocities of this HVC are in the range 103
to 354 \kms. We find emission from the disk up to LSR velocities of 170
\kms~ and restrict therefore the velocity range accordingly. The HVC
emission lines are very narrow at all velocities 
(Fig. \ref{FigEP_vau-sig}). The first moment of the column density
weighted distribution of the velocity dispersions is 11.5 \kms. There
are no obvious shifts between the center velocities of secondary and
primary components.
 
\subsection{Complex ACHV \& ACVHV}

\begin{figure}[!ht]
   \centering
   \includegraphics[width=9cm]{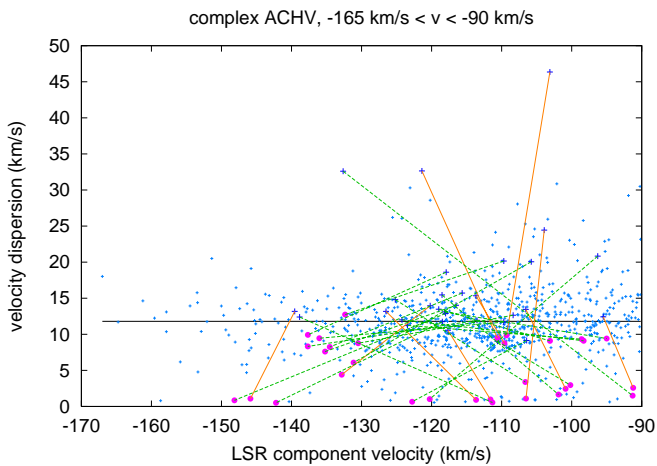}
\caption{Same as Fig. \ref{FigA_vau-sig} but for complex ACHV. }
\label{FigACHV_vau-sig}
   \end{figure}

First observations of the anti-center HVCs have been published by
\citet{Hulsbosch1968}, a more detailed discussion is given by
\citet{WvW}.  Neither distances nor metallicities are known.  4\% of the
observed positions show a multi-component structure. The lines are in
general broad in complex ACHV, even for most of the secondary
components. Fig.  \ref{FigACHV_vau-sig} shows the distribution of
component velocity dispersions. The velocity shifts show a broad
scatter, some of the components tend to have shifts of $\sim -30$ \kms.

\begin{figure}[!ht]
   \centering
   \includegraphics[width=9cm]{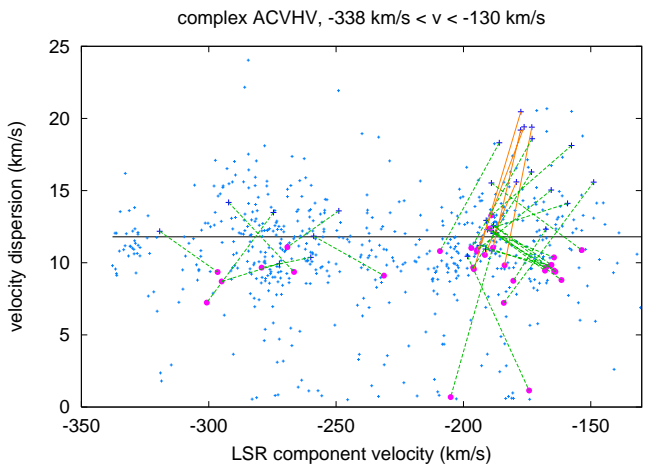}
\caption{Same as Fig. \ref{FigA_vau-sig} but for complex ACVHV. }
\label{FigACVHV_vau-sig}
   \end{figure}

Fig.  \ref{FigACVHV_vau-sig} shows a very different distribution of
component column densities for comples ACVHV. Secondary components are
not very frequent. A grouping at velocities $-200 \la v \la -160$
\kms~ for HVC ACI and $-300 \la v \la -250$ \kms~ for HVC 168--43--280 is
obvious.

\subsection{Complex WB}

\citet{Thom} report a distance determination in direction to this
complex and suggest a distance $ 7.7 < d < 8.8 $ kpc. Hence WB would
be located at $z \sim 7$ kpc and $ R \sim 12 $ kpc. As pointed out by
the referee, it may be questioned whether this limit applies for
complex WB. \citet{Thom} find an upper limit to a
small HVC (cloud 35) that was not included as part of this complex by
WvW. The "real" complex WB is a mostly-connected cloud at somewhat
lower latitudes.  There are many tiny positive-velocity HVCs at
somewhat higher latitudes in this part of the sky, which may be part
of the same complex. 

\begin{figure}[!ht]
   \centering
   \includegraphics[width=9cm]{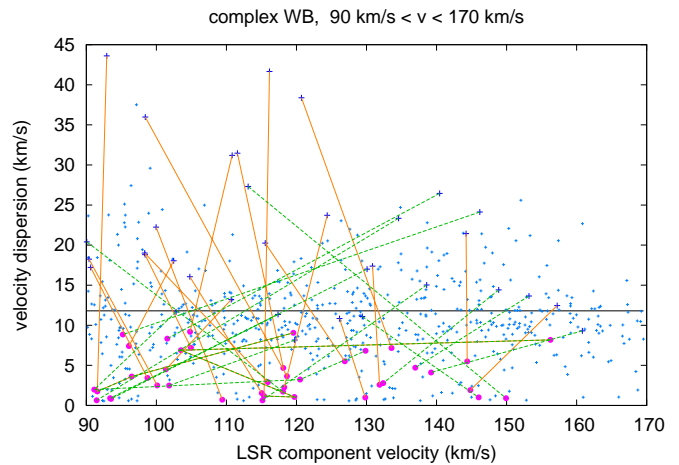}
\caption{Same as Fig. \ref{FigA_vau-sig} but for complex WB. }
\label{FigWB_vau-sig}
   \end{figure}

Most of the subsonic secondary components for complex WB (solid lines in
Fig. \ref{FigWB_vau-sig}) are shifted to higher velocities. Some of
the supersonic cores behave in the opposite way (dashed lines). 

\subsection{Complex WD}

\begin{figure}[!ht]
   \centering
   \includegraphics[width=9cm]{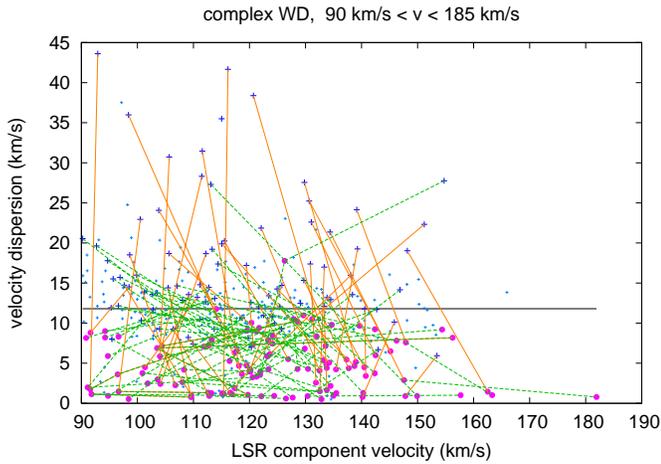}
\caption{Same as Fig. \ref{FigA_vau-sig} but for complex WD. }
\label{FigWD_vau-sig}
   \end{figure}

\citet{Wakker2004} lists this HVC with velocities $ 94 < v < 278 $ \kms. 
We find very little gas at $v \ga 170$ \kms. Secondary components are
shifted predominantly to higher velocities (Fig. \ref{FigWD_vau-sig}). 

\subsection{Complex WE}

\begin{figure}[!ht]
   \centering
   \includegraphics[width=9cm]{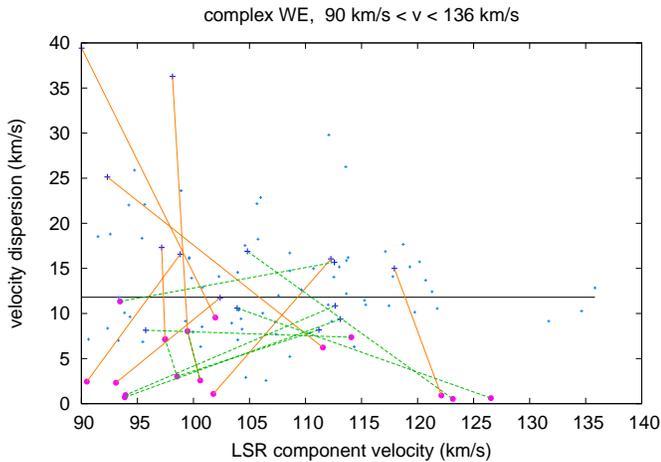}
\caption{Same as Fig. \ref{FigA_vau-sig} but for complex WE. }
\label{FigWE_vau-sig}
   \end{figure}

\citet{Sembach1991,Sembach1995} determined an upper distance limit of
$\sim 12.8$ kpc.  Similar to Complex WD we find no \hi emission at $ v
\ga 136 $ \kms. From \citep{Wakker2004} we expect emission up to $ v  =
195 $ \kms, but apparently this part of the HVC emission is weak.  There
are no systematic shifts in velocities between primary and secondary
components. Fig \ref{FigWE_vau-sig} looks rather chaotic.

\subsection{Complex R}

\begin{figure}[!ht]
   \centering
   \includegraphics[width=9cm]{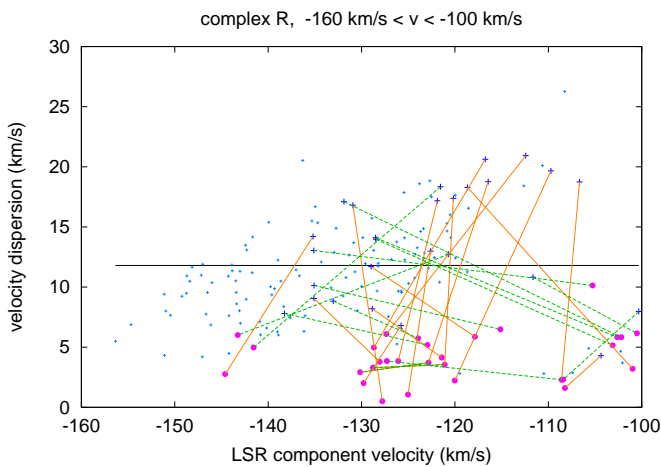}
\caption{Same as Fig. \ref{FigA_vau-sig} but for complex R. }
\label{FigR_vau-sig}
   \end{figure}

This HVC was first studied by \citet{Kepner1970}, who suggested that it
might be due to gas raining down to the disk.  At velocities 
$v \ga -120 $ \kms~ there is overlap with disk emission. Narrow
components in this velocity range may indicate interaction with the
disk (Fig. \ref{FigR_vau-sig}). 

\subsection{Complex G}

\begin{figure}[!ht]
   \centering
   \includegraphics[width=9cm]{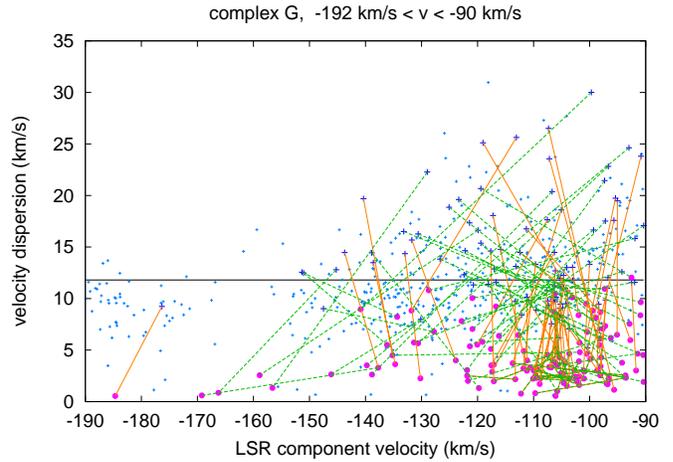}
\caption{Same as Fig. \ref{FigA_vau-sig} but for complex G. }
\label{FigG_vau-sig}
   \end{figure}

This gas has a lower distance limit of 1.3 kpc \citep{Wakker2001}. 
At velocities $v \la -160$ \kms~ we find no secondary components 
(Fig. \ref{FigG_vau-sig}).  

\subsection{Disk material}

\begin{figure}[!ht]
   \centering
   \includegraphics[width=9cm]{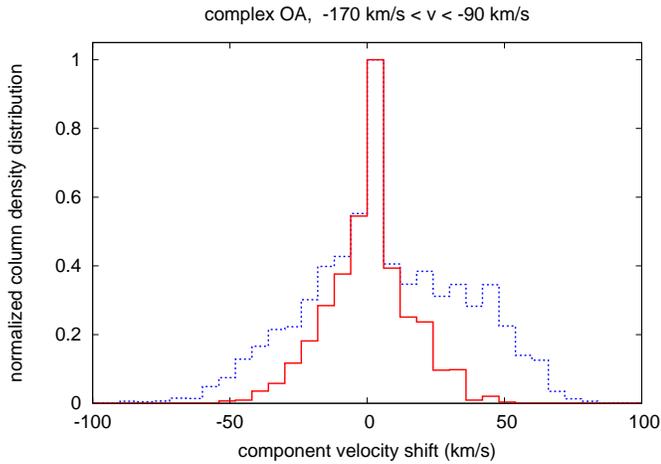}
\caption{Column density distribution of secondary components 
with respect to their velocity shifts $v_i-v_1$. 
Solid lines are for subsonic components with $|v_i-v_1| < \sigma_1$, dotted
lines for supersonic components up to Mach 2.35. }
\label{FigOA_histo_V}
   \end{figure}

Complex OA is usually interpreted as part of the outer arm, reaching
intermediate latitudes \citep{Habing,Haud1992}. From the rotation curve
we estimate a distance of 15 - 20 kpc for gas at velocities of $-130$
\kms. Secondary components are found at 42\% of all positions, twice the
average for all HVCs, but these components contain only 24\% of the
observed column densities (see Table 1). This gas has most probably
properties similar to the \hi in the disk. For such a multi-phase medium
one would expect random internal motions. Indeed, we find no
preferences, grouping or other relations between primary and secondary
components. Fig. \ref{FigOA_histo_V} reflects this situation. The
distribution of component velocity shifts is nearly symmetric.

\begin{figure}[!ht]
   \centering
   \includegraphics[width=9cm]{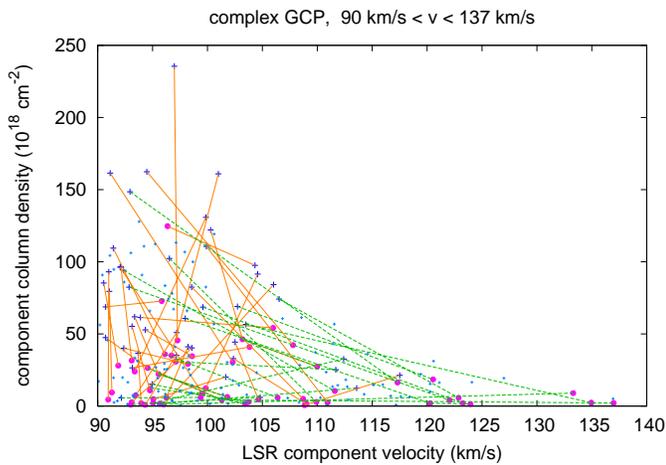}
\caption{Same as Fig. \ref{FigA_vau-NH} but for complex GCP. }
\label{FigGCP_vau-NH}
   \end{figure}

Similar to complex OA the number of positions in complex GCP with
multi-component structure is large, however the associated column
density is less prominent.  The relations between primary and secondary
components appear to be purely statistical
(Fig. \ref{FigGCP_vau-NH}). Some similarities to complex OA may be a
hint that this complex is closely associated with the disk.

\subsection{Non-detections or questionable cases}

\begin{figure}[!ht]
   \centering
   \includegraphics[width=9cm]{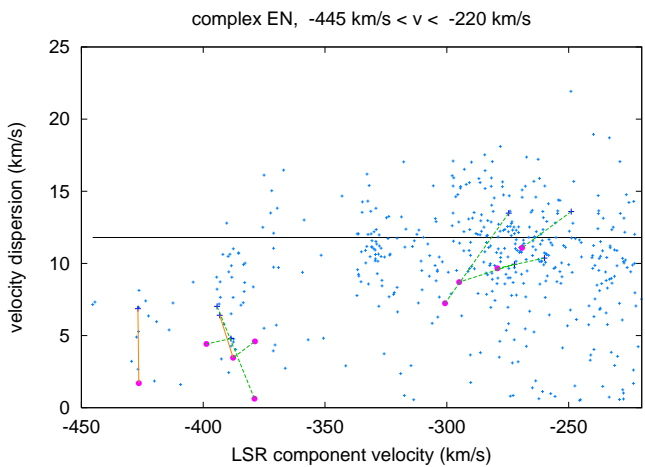}
\caption{Same as Fig. \ref{FigA_vau-sig} but for complex EN. }
\label{FigEN_vau-sig}
   \end{figure}

The extreme negative velocity clouds in complex EN contain only a few
multi-component detections (Fig. \ref{FigEN_vau-sig}).
\citet{Wakker2004} classifies these clouds with velocities $-465 < v <
-132 $ \kms. For $ v \ga -220$ \kms~ we found significant contributions
from Galactic plane gas.  We therefore excluded this range.

\begin{figure}[!ht]
   \centering
   \includegraphics[width=9cm]{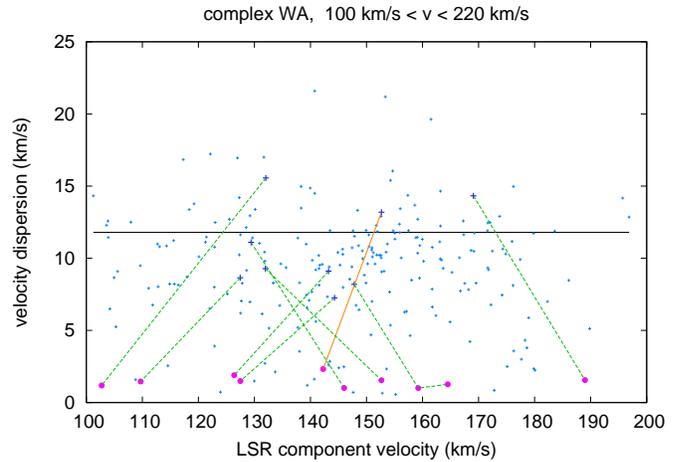}
\caption{Same as Fig. \ref{FigA_vau-sig} but for complex WA. }
\label{FigWA_vau-sig}
   \end{figure}

Complexes WA and WC contain very little gas in secondary
components. Most of the narrow lines have very low column densities and
may therefore be spurious.  WA and WC have each only two significant
secondary components.  Remarkable is, however, that most of the
secondary components in complex WA, if real, are very cold and shifted
by 36 \kms~ relative to the LSR velocity of the primary components
(Fig. \ref{FigWA_vau-sig}). WC is different, the components are also
cold but it shows a broad scatter of the velocity shifts.

In three complexes we found only one position with multi-component
structure, L, GCN, and D.  In complex P we have 5 detections, but 3 with
very low column densities. The complexes discussed in this
section have less than 3\% of the gas in cores. 

\section{HVC - IVC interaction}

Observational indications for interactions of HVCs with the Galactic
disk have been claimed for several HVC complexes, we refer to the most
recent review by \citet{Bruens2004}. HVCs approaching the disk are
expected to be decelerated by ram pressure \citep{Benjamin1997}.  In
turn, shocks may lead to fragmentation and phase
transitions. Head-tail structures \citep{Bruens2000} may develop and
velocity-bridges \citep{Pietz1996} may connect individual \hi HVC
fragments in the velocity-position space. Strong interaction may lead
to collisional ionization, observable in \ha~ \citep{Weiner1996} or
\ion{O}{vi} absorption \citep{Sembach2003}. X-ray
enhancements \citep{Kerp1996} or even $\gamma$-ray emission
\citep{Blom1997} may be caused.

\begin{figure}[!ht]
   \centering
   \includegraphics[width=9cm]{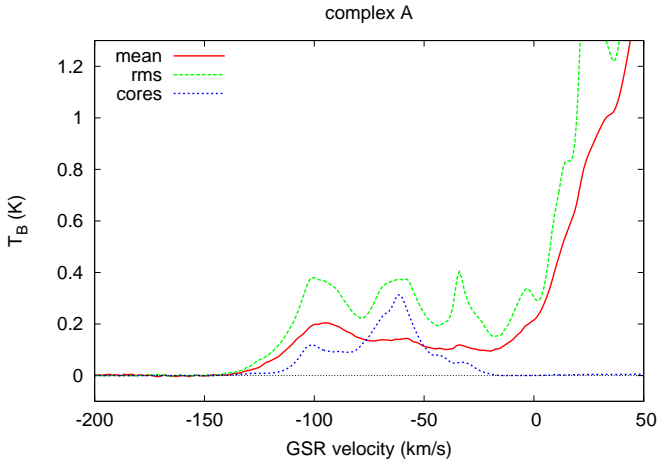}
\caption{Mean emission (solid line, red) and rms fluctuations (dashed
line, green) from the LAB survey calculated for all positions where HVC
emission was detected.  The lower dotted line (blue) shows the mean
emission from the secondary Gaussian components. The data are the same
as displayed in Fig. \ref{FigA_rms} but here we use GSR velocities. }
\label{FigAA_rms}
   \end{figure}

Here we intend to discuss whether from the Gaussian decomposition any
evidence for an interaction between HVCs and IVCs can be found.  The
spatial distribution of HVC complexes overlaps partly with IVC
features, see e.g. the reviews by \citet{Wakker2004} or
\citep{Albert2004}. The question arises whether the positional
agreement between both components is accidental or whether there might
be some evidence for an interaction.

\subsection{The HVC rest frame: LSR, GSR, LGSR or deviation velocity?} 

Up to now we avoided any interpretation about origin and distances of
HVCs. All parameters have been described in the Local Standard of Rest
(LSR) frame. Observations in this frame are corrected for the basic
solar motion and tied locally to a rotating Galactic disk. If HVCs are
associated with the Milky Way but not co-rotating with the disk, the
Galactic Standard of Rest (GSR) would be more appropriate for a
description. It is also possible that HVCs are building blocks from
which the Local Group was assembled \citep{Blitz1999}, in this case
the Local Group Standard of Rest (LGSR) would be the best
choice. Equations for a conversion between different systems are given
by \citet{deHeij2002c}.  Last, not least, HVCs are characterized
sometimes by their deviation velocities \citep{Wakker1991}. This
measure is not tied to an interpretation but allows to judge how
significant HVCs are separated from the Galactic disk emission. A
display of the HVC sky in deviation velocities as well as GSR
velocities is given by \citet{Wakker2004}.

Fig. \ref{FigA_rms} shows in the LSR rest frame the mean emission and
emission fluctuations derived for all positions of complex A that show
HVC emission. In addition we plot the mean emission for the cores
only. Most of the gas, in particular the emission from the cores, is
clearly separated from the disk and the IVC emission at $v \sim -50$
\kms. The picture changes significantly if we use GSR velocities
(Fig. \ref{FigAA_rms}). The HVC separation from the disk emission is
less obvious. Projection effects cause the core emission to be much
closer to the mean disk emission at $v \ga 0$ \kms. The prominent IVC
feature is washed out completely in velocity.  A projection of complex
A into the LGSR system shows a similar picture, shifted approximately
by 40 \kms, and we conclude that complex A, considered as a single
coherent cloud, is distinguished best from the local gas and the IVC
emission in the LSR system. IVCs, anyway, are best described in the
LSR system.

There is no space to display other examples in the GSR or LGSR frame
but it is clear from this example that the LSR frame is appropriate to
discuss the question whether some of the HVCs might be interacting
with the disk. In this frame the HVC velocities differ typically by
several tens of \kms~ from IVCs and the disk emission. An interaction
between HVC and disk gas necessarily would lead to strong shocks  and
we would expect to find cold, dense \hi clouds as tracers of such an
interaction.  Secondary components at low velocities, associated
with primary components at large velocities, may indicate such
shocks. The search algorithm used in Sect. 4 was biased since we
rejected any Gaussian component with velocities below the pre-defined
lower velocity limit (see column 4 of Table 1, typically $|v| > 90 $
\kms). 

To find features indicating a possible HVC/IVC interaction we extend
our search algorithm. We allow primary components at high velocities
to be associated with secondary components at low velocities. The mere
existence of such pairs of Gaussians is not a sufficient proof for an
interaction. Such components may show up accidentally along the line
of sight. However, finding numerous cases would be in favor for an
interaction. The interaction hypothesis can be rejected if our search
fails to find such component pairs. We discuss the easy case first.

\subsection{No indications for HVC/IVC interaction in complex 
A, MS, ACVHV, EN, WA, \& P}

Fig. \ref{FigA_rms} shows that along the line of sight to Complex A
also a significant contribution at intermediate velocities is
observed. Both, the mean emission profile as well as the rms deviation
from the mean, peak at $ v \sim -55$ \kms. However, from the Gaussian
decomposition, there is no significant evidence for secondary
components at velocities $v \ga -90$ \kms. Extending our search
algorithm does not disclose new components and we conclude that there
are no indications for IVC cores associated with primary HVC
components.

Our conclusion that there is no HVC/IVC interaction in direction to
complex A is consistent with previous distance determinations. Complex
A is at a z height of 5 to 7 kpc \citep{vanWoerden1999,vanWoerden2004},
far above the IVC gas layer with typical upper limits of $ 1 \la d \la 4$
kpc \citep{Albert2004}. No interaction is expected. 

Similar to complex A we find no indications for an HVC/IVC interaction for
the Magellanic stream and complexes ACVHV, EN, WA, and P. We list in Table  
1 the fraction $f_{NIV}$ of the column density observed at intermediate
velocities, outside the HVC velocity range listed in column 4 of Table
1. For all of these complexes $f_{NIV} < 0.01$.

\begin{figure}[!ht]
   \centering
   \includegraphics[width=9cm]{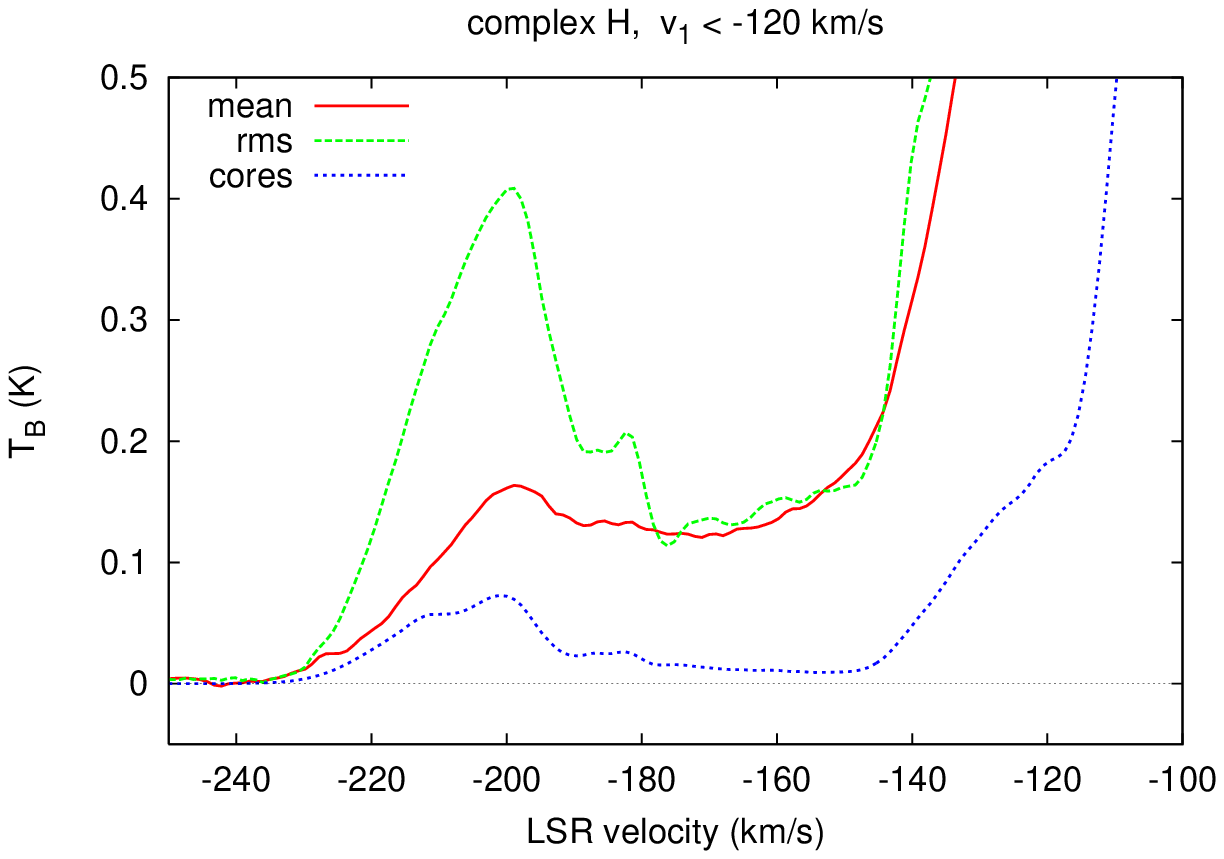}
\caption{Mean emission (solid line, red) and rms fluctuations (dashed
line, green) from the LAB survey calculated for all positions where HVC
emission was detected.  The lower dotted line (blue) shows the mean
emission from the secondary Gaussian components. }
\label{FigH_rms}
   \end{figure}

\subsection{Complex H}

Figs \ref{FigHH_vau-sig} to \ref{FigH_histo_V} in Sect. 4.3 gave
evidence for a bimodal distribution of the secondary components
associated with primary components in complex H. Cores at high
velocities move typically 20 to 30 \kms faster than the envelopes
while cores at low velocities are slow by 50 to 70 \kms. Fast cores
move predominantly subsonically, slow cores are mostly supersonically.
Fig. \ref{FigH_rms} shows the same bimodal distribution for the mean
emission (solid line) and its rms fluctuations (dashed). A broad
minimum exists between $ -180 \la v \la -150$ \kms, a ``velocity
bridge'' according to \citep{Pietz1996}, indicating a possible
interaction between gas at high and low velocities. The mean emission
from the cores (dotted line in Fig. \ref{FigH_rms}) has a similar
velocity bridge between $ -180 \la v \la -150$ \kms and a strong
increase for $ v \ga -150$ \kms. There is a marked contrast to complex
A (Fig. \ref{FigA_rms}) and the complexes discussed in the previous
section. We find a large number of narrow lines for $ v \ga -150$
\kms, exclude therefore chance coincidences and interpret the
secondary components as a population of cold, shocked clumps, most
probably caused by an interaction between HVC and disk gas. More than 
30\% of the total observed column density in direction to complex
H is at velocities $v > -120$ \kms.

A part of complex H was observed recently by \citet{Lockman2003} with
the GBT. Lockman suggested that this object could be a satellite of
the Milky Way, interacting with the Galactic disk. He determined a
distance of $d \sim 27$ kpc, or $R \sim 33 \pm 9$ kpc. The GBT
observations gave also evidence for a multi-phase medium. A group of
condensations at galactic coordinates $l \sim 131\deg$, $b \sim 1\deg$
is particularly interesting since its small scale structure at scales
of 4\arcmin~ is known \citep{Wakker1991S} and, most important, a spin
temperature of $T \sim 50$ K \citep{Wakker1991V} was measured.  This
allows for an accurate determination of an internal pressure $P/k \sim
100$ cm$^{-3}$K for the cores at a LSR velocity $v \sim -200$
\kms. Moreover, a similar pressure was determined with the GBT for the
surrounding envelope, a convincing argument for a multi-phase medium
in pressure equilibrium \citep{Lockman2003}.

How about the Milky Way disk? Using a model of the Galactic \hi
distribution \citep{Kalberla2003}, we expect an average pressure of
$P/k \sim 100$ cm$^{-3}$K for gas in the disk at a LSR velocity of $v
\sim -130$ \kms~ and a distance $d \sim 26$ kpc, or $R \sim 32$ kpc,
just at the position of complex H as estimated by
\citet{Lockman2003}. Using alternatively a rotation curve according to
\citet{Brand1993} would result in a slightly different LSR velocity of
$v \sim -124$ \kms~ and a 30\% larger pressure. An inspection of the
LAB survey gives a clear evidence for continuous \hi emission at such
velocities as expected for the outskirts of the disk, see also
\citet[Fig. 3 of][]{Blitz1999} or \citet[Fig. 3
of][]{Lockman2003}. Most of the emission at LSR velocities of $ -120
\la v \la -90$ \kms~ originates from the Galactic disk. Emission from
the clumps at $ -150 \la v \la -120$ \kms (Fig. \ref{FigH_rms}) is
most probably caused by a HVC/IVC interaction.

\begin{figure}[!ht]
   \centering
   \includegraphics[width=9cm]{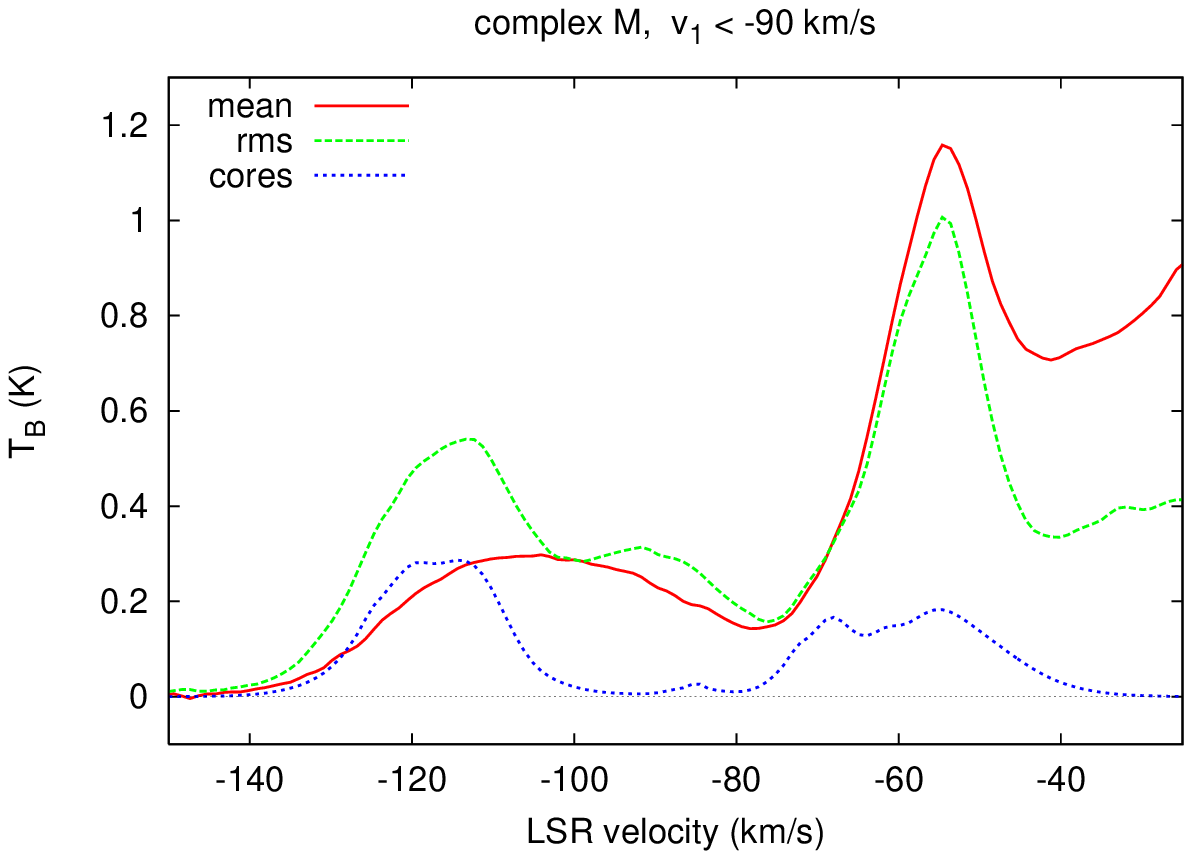}
\caption{Mean emission (solid line,red ) and rms fluctuations (dashed
line, green) from the LAB survey calculated for all positions where HVC
emission was detected.  The lower dotted line (blue) shows the mean
emission from the secondary Gaussian components. }
\label{FigM_rms}
   \end{figure}

\begin{figure}[!ht]
   \centering
   \includegraphics[width=9cm]{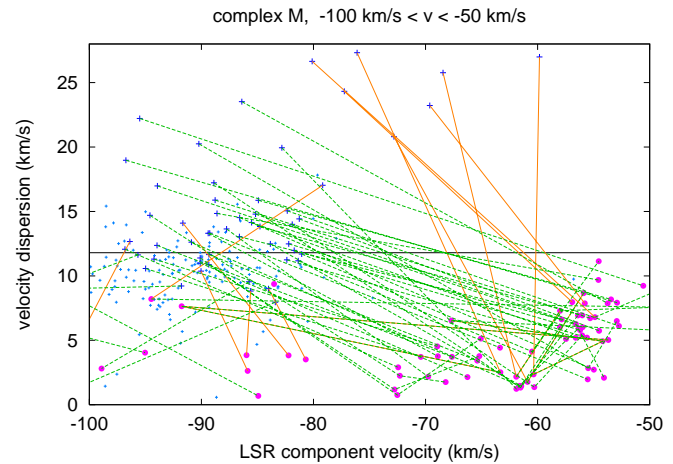}
\caption{Same as Fig. \ref{FigA_vau-sig} but for complex M. }
\label{FigM_vau-sig}
   \end{figure}

\subsection{Complex M}

We apply our search algorithm for IVC clumps also to complex M. We find
again a bimodal distribution of the core velocities. Clumps at $ -80 \la
v \la -40$ \kms (Fig. \ref{FigM_rms}) may be associated with HVC
envelopes at higher velocities. Fig. \ref{FigM_vau-sig} shows the
relation between cores and envelopes in more detail. We find a highly
non-random distribution, most of the lines connecting cores and
envelopes run parallel. The similarity with the low velocity part of
complex H (Fig. \ref{FigHH_vau-sig}) is striking.

Distance and metallicity of complex M were discussed recently by
\citet{vanWoerden2004}. The upper limit of its distance is $ z \la 3.5
$kpc. The high metallicity (0.8 solar) suggests an association with the
Intermediate-Velocity Arch, which lies at a distance of 0.8 to 1.8
kpc. The association between high velocity envelopes and low velocity
cores provides additional evidence for a possible interaction. 

\subsection{Complex C }

\begin{figure}[!ht]
   \centering
   \includegraphics[width=9cm]{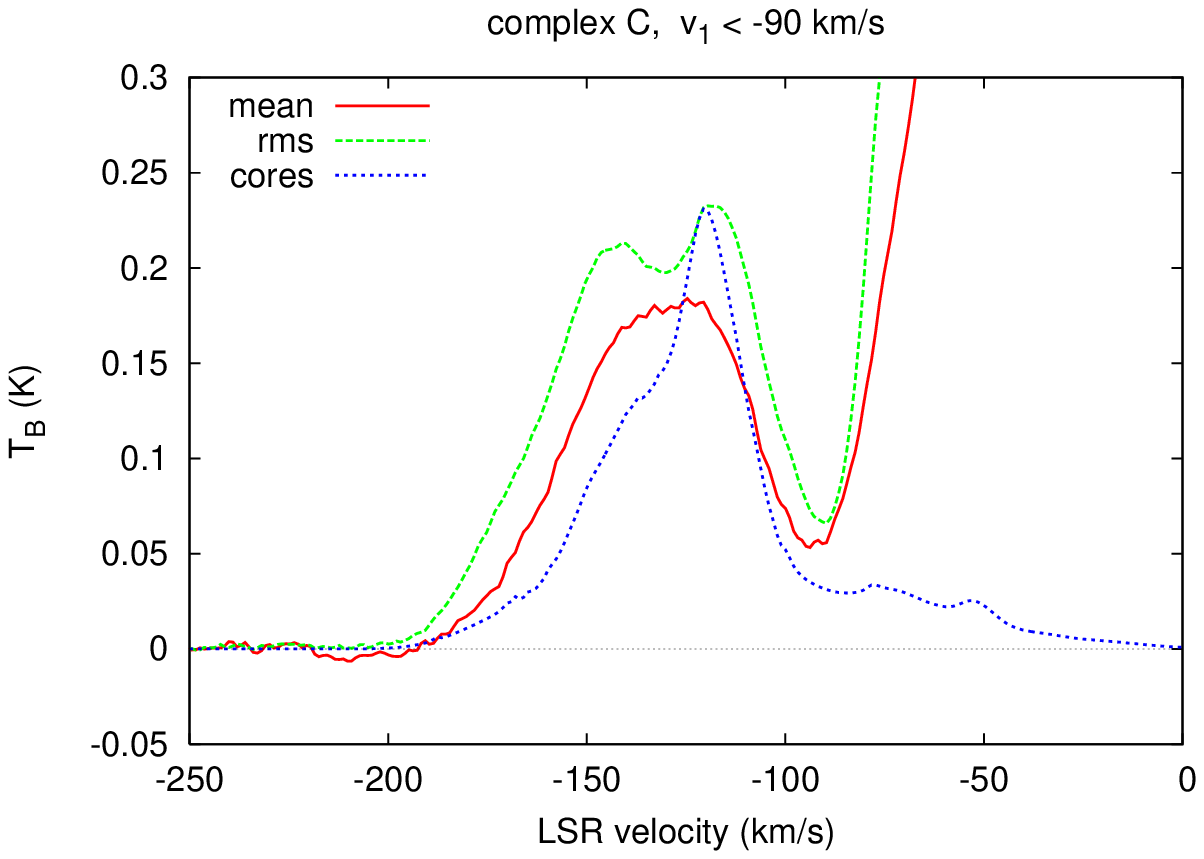}
\caption{Mean emission (solid line, red) and rms fluctuations (dashed
line, green) from the LAB survey calculated for all positions where HVC
emission was detected.  The lower dotted line (blue) shows the mean
emission from the secondary Gaussian components. }
\label{FigC_rms}
   \end{figure}

Also complex C is of particular interest. In Fig. \ref{FigC_rms} we
display profiles for the mean emission, its rms fluctuations and the
emission due to secondary components. The emission from the cores
peaks at $v \sim -120$ \kms and has an extended tail to lower
velocities. A number of secondary components appear to be related to
primary components at larger velocities, however the total amount of
gas in cores at intermediate velocities is lower than in the cases
discussed in Sect. 5.3 \& 5.4. Complex C is the second largest after
complex OA but apparently only a fraction of it shows signs of
interaction. 

Using the LDS, \citet{Pietz1996} detected \hi velocity bridges in complex C
connecting HVC gas with intermediate velocities. Effelsberg observations
with better resolution confirmed the reality of these features. Using
the ROSAT all-sky survey, \citet{Kerp1996} found soft X-ray enhancements
in the same directions and took this as an indication for an impact of
HVCs onto the Galactic disk. \citep{Bruens2000} found head-tail
structures and concluded that these cometary features are
caused by interaction. \citet{Wakker2004} argues against a possible
connection between complex C and intermediate velocity clouds (IVC) in
the intermediate velocity Arch IV \citep{Kuntz1996}.  Abundances and
their variations in direction to complex C may provide clues for the
origin of this HVC gas.  \citet{Wakker1999,Richter2001a,Tripp2003} found
evidence for the infall of low-metallicity gas onto the Milky Way, we
refer to the most recent discussion by 
\citet[][Sect. 6.1]{vanWoerden2004}.

\subsection{Complex WB, WD, WE, WC, R, G, GCP, \& OA }

\begin{figure}[!ht]
   \centering
   \includegraphics[width=9cm]{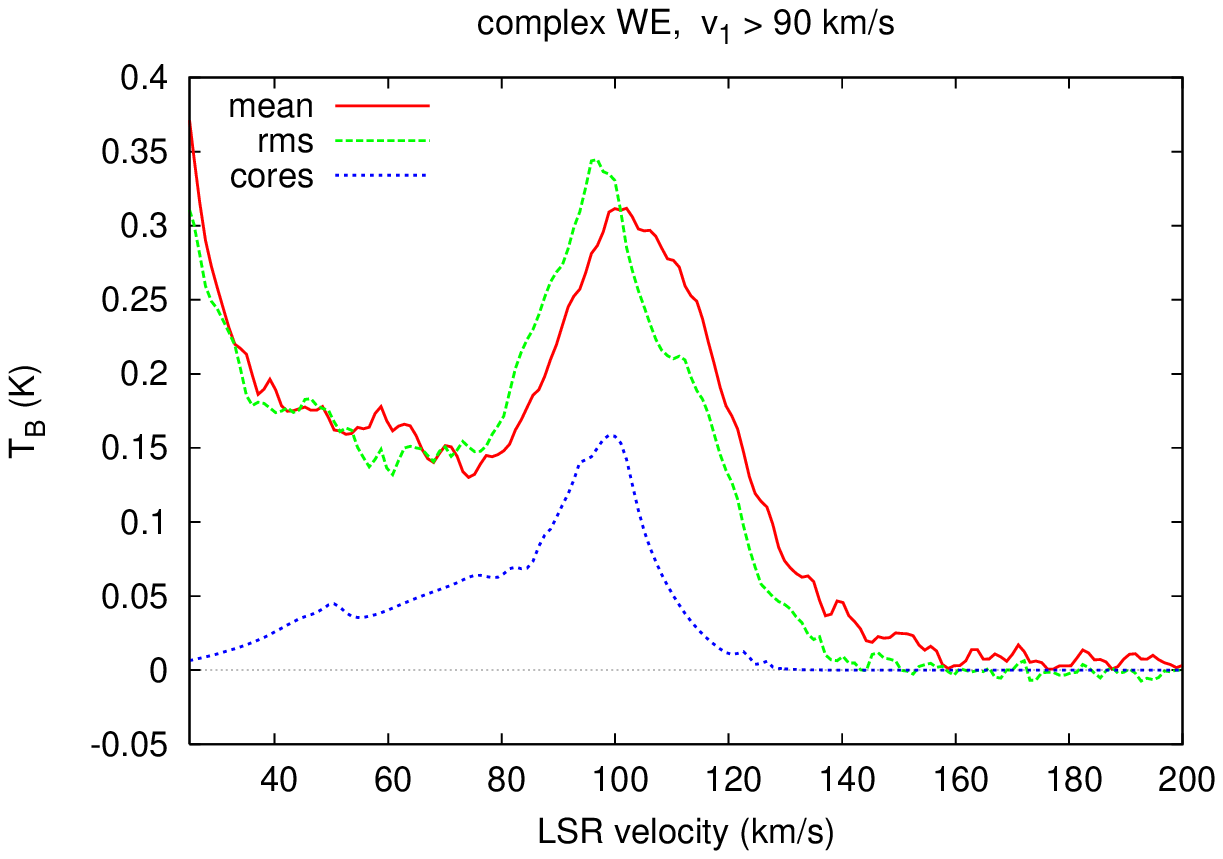}
\caption{Mean emission (solid line, red) and rms fluctuations (dashed
line, green) from the LAB survey calculated for all positions where HVC
emission was detected.  The lower dotted line (blue) shows the mean
emission from the secondary Gaussian components. }
\label{FigWE_rms}
   \end{figure}

In all cases we find low velocity wings for the mean emission of the
secondary components. The mean emission profiles in direction to these
HVCs, similar the rms deviations, have extended wings toward zero
velocities. The wings for the mean emission are caused by differential
Galactic rotation. In Sect. 5.2 we found no indications that secondary
components may be seriously affected due to blending with foreground or
background sources. For the complexes discussed here the lines of sight
are more complex and we cannot exclude that a significant part of the
secondary components at intermediate velocities may be spurious.

Complex WE is of particular interest since an upper
distance limit of $ d \la 12.8$ kpc ($z \la 3.2$kpc) is known
\citep{Sembach1991,Sembach1995,Wakker2001}. Fig. \ref{FigWE_rms}
displays the derived spectra for mean emission, rms deviations, and
emission from cores. 


\section{Discussion}

We have decomposed the LAB survey database into Gaussian components.
All HVC complexes have been searched for indications of a multi-phase
medium. We obtained clear indications for two \hi phases, broad lines
with a typical velocity dispersion of 12 \kms~ dominate. Narrow lines
with typical dispersion of 3 \kms~ are found predominantly in
regions with high HVC column densities. In many cases the secondary
components have systematic velocity shifts relative to the broad primary
components but we observe also a large scatter. The properties of the
HVC cores, listed so far, are known since \citet{Cram1976}.  However,
our investigations provide a much larger sample. All of the major HVC
complexes show indications for a multi-component structure.
Depending on the individual complexes up to 38\%, on average 24\%, of the
positions with HVC emission have multiple HVC Gaussian
components. These are predominantly positions with large column
densities.  Up to 27\%, on average 20\%, of the observed HVC flux
originates from cold cores. Our analysis covers large complexes only.  In
the case of compact HVCs (CHVCs) \citet{deHeij2002a} find 4 -- 16\% in
the cold phase. 

Clear indications for a multi-phase structure in complex EN, WA, WC,
P, L, GCN, and D are missing. All-together these cover about 6\% of
of all sky positions where we find HVC emission. The question arises whether
these complexes might be special or whether the LAB data simply might
not be adequate for these complexes. \citet{Giovanelli1973} have
pointed out that cores are small scale features, barely resolved with
a beam of 0.5\deg. The LAB has a similar beam size and we need to take
into account that this fact causes some restrictions. 

The distances to
most of the HVCs are unknown but complex H, MS, and EP are important
exceptions. Cold cores do exist at a distance of $\sim 50$ kpc. This
implies that the cores must have typical dimensions of a few hundred
pc, otherwise beam dilution would heavily degrade the signal. High
resolution observations are needed to resolve the cores and in
particular to verify whether there is an interaction between gas at
high and intermediate velocities
\citep{Pietz1996,Bruens2000,Lockman2003}.

\citet{Wakker1991S} have demonstrated how important
high-resolution interferometric observations are for a detection of
cores. They used the WSRT to map two clouds. Six compact isolated HVCs
have been mapped by \citet{Braun2000} and six more by
\citet{deHeij2002a}. \citet{Burton2001}, using the Arecibo telescope,
confirm that essentially all CHVCs have a core-halo structure, for
discussion see Sect. 5 of \citet{Burton2004}. Velocity dispersions
derived for CHVCs are somewhat smaller, but taking beam smearing into
account, in any case consistent with values derived by us. We conclude
that most probably our results can be generalized in the sense that
{\it all } HVCs have a multi-phase structure. The probability to
detect cold cores would then depend merely on telescope sensitivity
and resolution.

The conditions for a multi-phase medium in the Galactic disk and the
halo have been discussed by \citet{Wolfire1995a,Wolfire1995b}. A stable
two-phase \hi gas can exist over a range of heights but only within a
narrow range of pressures at each height. It was shown that a hot ($T
\sim 1 - 2 \cdot 10^6$ K) halo can provide the necessary
pressure. Observations of a multi-phase medium can, in principle,
constrain distances of HVCs, provided their metallicity and dust-to-gas
ratio is known.

For a few complexes a multi-component structure is rather
unexpected.  The first is the part of the Magellanic stream at LSR
velocities $ 50 \la v \la 362 $ \kms ($ 50 \la v \la 201$ \kms without
the interface region). This gas is approximately at a distance of 50
kpc. \citet{Wolfire1995b} predicted that a multi-phase medium
associated with the Magellanic stream should exist only at distances
$z \la 20$ kpc above the disk.  Fig. \ref{FigMS+_vau-sig}, however,
shows a well defined two-component structure. The number of positions
with such a structure as well as the associated column densities are
even larger than the average, see Table 1. Cores with narrow lines
were found also in the leading arm (complex EP). This HVC is usually
estimated to be at a similar distance, a two-phase \hi medium is
therefore also unexpected. High resolution Parkes observations of the
Magellanic stream and the leading arm confirm a multi-component
structure for the leading arm \citep{Bruens2005}.  Molecular hydrogen,
detected by \citet{Richter2001b} and \citet{Sembach2001}, is another
indicator for a multi-phase structure.

The trailing part of the Magellanic stream at negative LSR velocities
differs significantly.  It was found to have less secondary
components. This may indicate a lower confining halo gas pressure
and/or a larger distance for this part, as expected for a tidal origin
\citep{Gardiner1996}.

Comparing the observational evidence for a multi-phase medium with
model predictions \citep{Wolfire1995b}, we find it difficult to
explain such a medium, at least for an assumed temperature of
$T \sim 10^6$ K. Using the model of the gas distribution proposed by
\citet{Kalberla2003} we estimate a pressure of $P/k = nT \sim 20-40~
{\rm cm}^{-3}~ {\rm K}$ for complex MS+ and EP, probably still too low
to explain the existence of a multi-phase medium. It appears necessary
to reconsider the conditions that allow a multi-phase \hi medium at
such a distance in more detail.

The most recent analysis on the temperature of the hot halo is by
\citet{Pradas2005} who have correlated the LAB survey with the ROSAT
all-sky survey. Using simultaneously all ROSAT energy bands they were
able to fit the observations with two plasma components, the first
originating from the local hot bubble with a temperature of $T =
10^{5.9}$ K and the second from the halo with a temperature of $T =
10^{6.15}$ K. This increases estimates for a lower limit of the halo
pressure by a factor of two.  Additional arguments for a high pressure
in the halo have been given by
\citet{Weiner1996,Sembach2003,Fox2004,Fox2005}. Halo densities of up
to $10^{-4}$ cm$^{-3}$ are proposed close to the Magellanic stream and
the leading arm. Such densities would increase the halo pressure up to
a factor 10. Interestingly, $10^{-4}$ cm$^{-3}$ is also the density
required to explain compression fronts and head-tail shaped features
in HVCs \citep{Quilis2001}, that have been found by
\citet{Bruens2000,Bruens2001} in almost all of the HVC complexes.

Another unexpected case is complex H, lying in the plane of the Galactic
disk. \citet{Lockman2003} located this HVC at a distance $R \sim 33$ kpc
from the center and found evidence for a multi-phase medium at a
pressure of $P/k = nT \sim 100~ {\rm cm}^{-3} {\rm K}$. In Sect. 5 we
argued that the gas in the disk has a similar pressure. Also the disk
shows a two-component structure and there are indications for an
interaction between disk and HVC. According to \citet{Wolfire2003} a
multi-phase medium may exist in the Milky Way at distances $ 3 \la R \la
18$ kpc. It appears necessary to consider the question whether 
the upper distance limit needs to be revised. Our results indicate that 
current models underestimate the halo gas pressure.

The situation may be different in compact high velocity clouds
(CHVCs). \citet{Sternberg2002} discussed the conditions for a
multi-phase structure in such objects. CHVCs have not been included in
our sample since the spatial resolution of the LAB survey is not
appropriate for these objects. 

HVC complexes show some similarities in their internal velocity
structure. To characterize this we derived velocity shifts $v_i - v_1$
between cores and envelopes and characterized the average internal
motions by the first and second moments of the column density weighted
distribution of the velocity shifts. Most of the HVCs have internal
turbulent motions with Mach numbers between 1 and 2.  These internal
motions within HVCs should be compared with the properties of the disk
gas.  \citet{Heiles2003}, from an analysis of Arecibo absorption lines,
find that cold clumps in the disk move only slightly supersonic within
the surrounding warm neutral medium. Cores within HVCs are apparently
faster. Our analysis may be biased. We excluded Gaussian components with
velocity shifts $v_i - v_1 > 2.35 \sigma_1$. Some HVC cores with very
large turbulent motions may have been rejected, the derived typical Mach
number $ M \sim 1.5$ is therefore a lower limit only. 

Next we consider the column density weighted standard deviation of
{\it Gaussian component center velocities}.  \citet{Heiles2003} find
$\sigma_{WNM} \sim 11$ \kms~ for the WNM and $\sigma_{CNM} \sim 7$
\kms~ for the CNM in the disk, consistent with previous results
\citep[e.g.][]{Mebold1982}.  The situation for the relative motion of
cold cores in HVCs is very different. We determine a column density
weighted velocity dispersion $\sigma \sim 20$ \kms~ for velocities
$v_{core}-v_{envelope}$ (last column in Table 1).  
\citet{Davies1976} have noted this effect first in complex A IV and
described it as feature-to-feature velocity differences. HVC cores
typically move three times faster than cold clumps in the disk, only
complex M has $\sigma \sim 10$ \kms~ and deviates significantly in
this respect.

The envelopes share the highly turbulent state of the clumps.
\citet[][their Table 2]{Blitz1999} determined the variation of HVC line
centroids from one position to the other and derived $\Delta v = 30$
\kms. This, once more, is three times the value derived by
\citet{Heiles2003} for the warm \hi phase in the disk. Individual HVC
components represent a highly turbulent gas phase. The turbulent energy
density of individual clumps of HVC gas is an order of magnitude larger
than that of comparable clumps in the Galactic disk.

\begin{figure}[!ht]
   \centering
   \includegraphics[width=9cm]{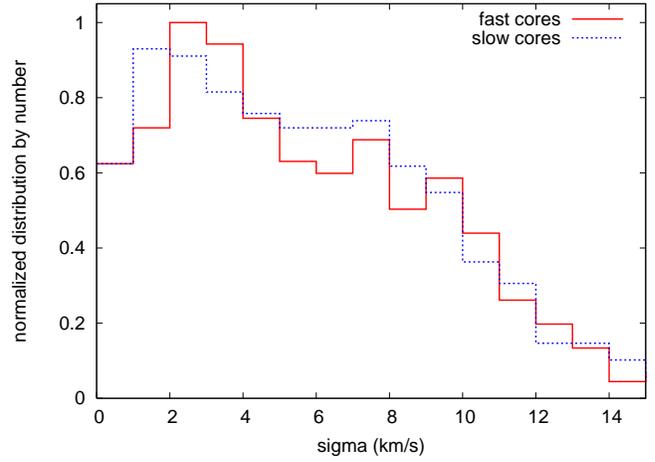}
\caption{Normalized frequency distribution of secondary component line
widths. The solid line represents secondary components moving faster
than the primary components, $|v_i| > |v_1|$, the dotted line those
moving slower.  Selected were all complexes except OA and all secondary
components with velocities $|v| > 90 $ \kms. }
         \label{Figclumpsn}
   \end{figure}

Despite having large internal motions, HVCs tend also to have ordered
internal motions relative to their envelopes. This can be seen best for
complex H in Fig. \ref{FigHH_vau-sig} for $ v \ga -160$ \kms.  Table 1
shows that cores are typically 5 \kms~ faster than the associated
envelopes. This is in good agreement with \citet{Cram1976} who derived
an average velocity difference of 4.5 \kms. Contrary, we found also a
population of clumps that move slower than the associated primary
components, best seen in Fig. \ref{FigHH_vau-sig} for $ v \la -160$
\kms. Is there a systematic difference between fast and slow moving
cores?

We calculated the distribution of velocity dispersions for the
secondary components of all HVC complexes, except complex OA. The
outer arm was excluded since most probably this complex belongs to the
disk population.  We distinguish between fast moving HVC cores ($|v_i|
> |v_1|$) and slow moving HVC cores ($|v_i| < |v_1|$) both with $|v| >
90 $ \kms. Fig. \ref{Figclumpsn} shows that components with $\sigma
\sim 2 $ \kms~ are most frequent but there is also a broad wing due to
components with larger dispersions. Both distributions are very
similar, systematical deviation probably barely significant. Internal
motions of cold cores within the envelopes of HVC complexes are on
average well balanced.

An inspection of Table 1 gives the impression that those HVC
complexes that have the lowest observed gas fraction in cores $f_n$
(by number) or $f_N$ (by column density) tend to have the most
negative velocities. We therefore calculated the number of positions
with secondary components relative to the total number of components
and the sum of column density in secondary components relative to the
total column density as a function of the component velocity. In both
cases we find a clear trend that the gas fraction in cores
increases systematically towards the positive velocities. This remains
valid also after repeating the analysis in the GSR or LGSR system.

\section{Summary and conclusion}

Using the LAB survey we derived the main observational parameters for
most of the well known HVC complexes. Performing a consistency check
with the WvW catalog \citep{WvW} we found a good agreement for the
peak temperatures. Center velocities of the WvW catalog turned out to
be accurate to 8 \kms but large discrepancies are found for the WvW
velocity dispersions. These are not an appropriate description of the
line widths and systematically biased. On average the WvW line widths
and also the derived column densities need to be scaled upward by a
factor of 1.4.

A systematic bias in the WvW catalog has some important
consequences. Most of the derived global properties for HVCs, like
masses of HVC complexes, also the net inflow of \hi HVC gas come from
this catalog \citep{WvW97}.  These numbers need to be revised upwards by
40\%. Pressures, derived for the diffuse part of the HVCs
\citep[e.g.][]{Wakker1991S} depend on $\sigma^2$, hence need to be
revised by a factor of $\sim 2$. Timescales estimated for a free
expansion of HVCs \citep{WvW97} decrease accordingly.

Our Gaussian decomposition of the LAB survey led to the result, that
most of the HVC complexes have a multi-phase medium. Cores with narrow
lines ($\sigma \ga 2$ \kms) are embedded in envelopes with broad lines
($\sigma \sim 12$ \kms). According to \citet{Wolfire1995b} this may be
taken as an indication that the HVCs are on average in pressure
equilibrium with a surrounding hot halo. We find a multi-phase structure
even at large distances, for complex H at $R \sim 33$ kpc, for the
Magellanic stream, and the leading arm at distances $D \sim 50$
kpc. Current models do not support a sufficiently large halo pressure at
such distances and need to be re-discussed.  The fact that more than 90\% of
the HVC gas in complexes, excluding CHVCs, has a multi-component
structure strengthens arguments that these clouds are associated with
the Milky Way.

We characterize the motion of HVC cores within the associated envelopes
by defining relative internal motions $v_{core}-v_{envelope}$ and Mach
numbers $M = |(v_{core}-v_{envelope})|/\sigma_{envelope}$. On average we
find Mach 1.5. Despite the fact that some HVC cores show ordered motions
there are on average no preferences for the direction of the internal
motions. The case $v_{core}-v_{envelope} > 0$ is as frequent as
$v_{core}-v_{envelope} < 0$. The center velocities for the cores within
a HVC complex show a typical dispersion of 20 \kms.  The relative
motion of HVC clumps is three times larger than the value derived for
the motion of comparable clumps of CNM within the Galactic disk. HVCs
are highly turbulent. The internal turbulent energy density of
individual clumps of HVC gas exceeds that of the disk gas by an order of
magnitude.

\begin{acknowledgements}
{We thank Leo Blitz for drawing our attention to inconsistencies
between the LDS and the WvW database, further for a machine readable
version of the WvW database and active support. T. Robishaw is
acknowledged for help with the database.  We are grateful to
W.B. Burton, J. Kerp, Ph. Richter, and T. Westmeier for valuable
suggestions and discussions. Last, not least, the referee is
acknowledged for constructive criticism. The participation of U. Haud
in this project was supported by the Estonian Science Foundation grant
no.  6106.}

\end{acknowledgements}

\onecolumn 
\begin{table}
\caption{Properties of HVC complexes. The columns are: name,
longitude, latitude, velocity range, number of positions observed
$n_{pos}$ , fraction of positions with cores: $f_n$, fraction of column
density in cores relative to the total column density: $f_N$, 
fraction of column density in cores at intermediat velocities: $f_{NIV}$,
first moment of the column density weighted distribution of velocity
dispersions: $ <\sigma_{NH}> $, first moment of the distribution of
velocity shifts $v_i - v_1$ for $M < 1$:  $ S(M<1) $, 
the same for $M < 2.35$: $ S(M<2.35)$, and second moment of the distribution of
velocity shifts $v_i - v_1$ for $M < 2.35$: $ \Delta S$ } 
\centering
\begin{tabular}{lccccccccccc}
\hline Name & Gal. long.  & Gal. lat.  & Vel. range & $ n_{pos} $ & 
$f_n$ & $f_N$ & $f_{NIV}$ & $<\sigma_{NH}>$ & 
S($M<1) $ &  S($M<2.35) $ & $ \Delta S$\\ 
{} & $[{\deg}]$ & $[{\deg}]$ & {[ \kms ]}  & {} & {} 
{} & {} & {} & {[ \kms ]} & {[ \kms ]} & {[ \kms ]} & {[ \kms ]}\\ 
\hline
all HVCs & & & $> |90|$ & 17336  & 0.238 & 0.208 & 0.207 & 11.8 & 
-1.5 & 0.5 & 16.3\\
all --OA & & & $> |90|$ & 10139  & 0.111 & 0.157 & 0.120 & 11.7 & 
-0.3 & 5.8 & 20.1\\
\hline
OA & 48.0 : 190.0 & -5.5 : 35.0 & -168.6 : -90. & 7197 & 0.416 &
0.229 & 0.226 & 11.8 & -1.8 & -0.8 & 14.8 \\
\hline
A & 130.0 : 172.8 & 22.0 : 49.0 & -216.8 : -94.0 & 727 & 0.129 &
0.114 & 0.003 & 11.1 & -3.7 & -7.5 & 12.6 \\
C & 40.5 : 147.6 & 17.5 : 61.0 & -202.2 : -90. & 2646 &  0.061 &
0.061 & 0.015 & 11.2 & -1.3 & 1.2 & 18.3 \\
H & 109.0 : 152.5 & -16.0 : 13.5 & -227.8 : -90. & 1284 & 0.255 &
0.401 & 0.208 & 11.3 & -0.6 & 16.3 & 21.9 \\
H(-120) &  &  & -227.8 : -120. & 1199 & 0.169 &
0.154 & 0.378 & 12.0 & -4.5 & 4.4 & 20.1 \\
M & 129.0 : 198.1 & 47.5 : 69.0 & -128.0 : -90. & 277 & 0.058 &
0.058 & 0.050 & 10.3 & -4.2 & -7.3 & 9.7 \\
MS+ & 281.9 : 333.4 & -84.5 : -61. & 50. : 200.8 & 150 & 0.380 &
0.270 & 0.000 & 13.4 & -2.7 & -4.4  & 19.1 \\
MS+W & 269.4 : 333.4 & -84.5 : -22.5 & 50. : 361.6 & 1109 & 0.548 &
0.563 & 0.000 & 13.3 & -1.6 & -1.1  & 22.6 \\
MS-- & 11.4 : 356.2 & -85.0 : -39.0 & -405.9 : -50. & 668 & 0.075 &
0.110 & 0.000 & 13.7 & -11.7 & -8.6 & 22.8 \\
EP & 244.0 : 312.5 & -27.0 : 28.0 & 170. : 317.4 & 290 &  0.238 &
0.135 & 0.002 & 10.6 & 0.9 & 1.7 & 14.8 \\
ACHV & 139.3 : 197.0 & -52.0 : -6.0 & -167.0 : -90. & 853 & 0.036 &
0.036 & 0.006 & 12.1 &  1.6 & -1.7 & 19.8 \\
ACVHV & 154.7 : 191.5 & -54.0 : -7.0 & -337.8 : -130.5 & 646 & 0.048 &
0.064 & 0.000 & 11.6 & -14.1 & -1.6 & 23.9 \\
WB & 224.5 : 268.8 & 0.5 : 45.5 & 90. :  169.6 & 599 & 0.057 &
0.041 & 0.119 & 12.6 &  5.5 & -0.9 & 20.7 \\
WD &  260.0 :   313.0 & 8.0 : 20.0 & 90. :  181.9 & 297 & 0.306 &
0.120 & 0.083 & 12.1 & 4.4 & 5.5 & 18.5 \\
WE &  291.5:  324.5 & -25.5 :-8.5  & 90. : 135.8 & 85  &  0.176 &
0.111 & 0.084 & 14.3 &  1.3 & -0.7 & 14.2 \\
R &   62.0 : 73.5  & 5.5 :15   & -156.3 : -100.3 & 142  &  0.190 &
0.096 & 0.325 & 11.6 & -2.7 & 1.6 & 15.0 \\
G &  79.0 : 121.5 & -19.0 : -1.0   & -189.9 : -90. & 521  &  0.190 &
0.138 & 0.298 & 12.2 & 0.8 & 1.8 & 16.5 \\
GCP & 34.5 :64.0  & -41.5 : -9.5  & 90. : 137.0 & 153 & 0.327 &
0.123 & 0.082 & 11.6 &  2.3 & 6.3 & 12.6 \\
EN & 29.0 : 187.5 & -59.5 : 33.5 & -445.4 : -220. & 481 & 0.017 & 0.015 &
0.000 & 11.1 & 2.9 & -7.0 & 15.4 \\
WA & 231.5 :  273.0 & 23.0 : 45.0 & 100. :  196.9 & 230 &  0.039 &
0.007 & 0.001 & 10.1 & -10.4 & -0.6 & 18.3 \\
WC & 208.5 : 259.0 & -33.0 : 3.5 & 90. :  212.7 & 156 & 0.071 &
0.028 & 0.433 & 13.4 & 7.2 & -5.3 & 16.5 \\
P & 107.0  : 132.5 & -39.0 : -30.0  & -429.4 : -320.4 & 67 & 0.060 & 0.048 &
0.000 & 8.5 & 3.0 & 5.1 & 9.4  \\
L & 340.5 : 348.5 & 31.0 : 41.5 & -156.0 : -90. & 30 &  - & - &
- & 12.3 & - & - & - \\
GCN & 0.5 : 49.5 & -40.0 : 10.0 & -334.6 : -177.0 & 125 & - & - & - & 
9.3 & - & - & - \\
D & 81.5 : 84.5 & 24.0 : 25.5 & -199.6 : -15.4 & 6 &  - & - &
- & 8.4 & - & - & - \\

\end{tabular}
\end{table}

\end{document}